\documentclass[aps,pre,preprint,nofootinbib]{revtex4}

\usepackage{amsmath,amssymb,amsfonts,graphicx,framed}
\usepackage{array,color}

\begin{document}

\title{A class of $2\times2$ correlated random-matrix models with Brody spacing distribution}

\author{Jamal Sakhr}
\affiliation{Department of Mathematics, The University of Western Ontario, London, Ontario, Canada}
\altaffiliation[Current Address: ]{Kinlin School, Fanshawe College, London, Ontario, Canada}

\date{\today}

\begin{abstract}
A class of $2\times2$ random-matrix models is introduced for which the Brody distribution is the exact eigenvalue spacing distribution. The matrix elements consist of constrained finite sums of an exponential random variable raised to various powers that depend on the Brody parameter. The random matrices introduced here differ from those of the Gaussian Orthogonal Ensemble (GOE) in three important ways: the matrix elements are not independent and identically distributed (i.e., not IID) nor Gaussian-distributed, and the matrices are not necessarily real and/or symmetric. The first two features arise from dropping the classical independence assumption, and the third feature stems from dropping the quantum-mechanical conditions imposed in the construction of the GOE. In particular, the hermiticity condition, which in the present class of models, is a sufficient but not necessary condition for the eigenvalues to be real, is not imposed. Consequently, complex non-Hermitian $2\times2$ random matrices with real or complex eigenvalues can also have spacing distributions that are intermediate between those of the Poisson and Wigner classes. Numerical examples are provided for different types of random matrices, including complex-symmetric matrices with real or complex-conjugate eigenvalues. Various generalizations and extensions are discussed including a simple modification that effectuates cross-over transitions between other classes of eigenvalue spacing statistics. The case of a cross-over transition between semi-Poisson and Ginibre spacing statistics is presented as a novel example. 
\end{abstract}

\maketitle

\section{Introduction}

Eigenvalue systems possessing nearest-neighbor spacing distributions (NNSDs) intermediate between those of the Poisson and Gaussian Orthogonal Ensemble (GOE) classes are ubiquitous in physics \cite{classi,Haake91,Graf92,Kudrolli94,Leitner97,Cheng02,Petit15,Roy17,Zhang19,Sierant19}. In the Poisson case, the exact NNSD is given by 
\begin{equation}\label{Eqn:Poisson}
    P_{P}(z) = \exp(-z).
\end{equation}
In the GOE case, the exact NNSD $P_{GOE}(z)$, which is sometimes called the Mehta-Gaudin distribution, cannot be expressed in closed-form (i.e., in terms of a finite number of elementary functions). An excellent analytical approximation to $P_{GOE}(z)$ is however given by the Wigner distribution \cite{Mehta,Haake}
\begin{equation}\label{Eqn:Wigner}
    P_{W}(z) = \frac{\pi}{2} z \exp\left( -\frac{\pi}{4} z^2 \right), 
\end{equation} 
which is actually the exact NNSD for a Gaussian ensemble of real symmetric $2\times2$ random matrices.
In the above equations, the variable $z$ is the standardized or mean-scaled spacing between consecutive eigenvalues.\footnote{To be precise, $z$ is the probability density variable corresponding to the random variable $Z\equiv {S/\mu_S}$, where $S$ denotes the nearest-neighbor spacing (a random variable) and $\mu_S$ denotes the mean of $S$ (a constant).} 

Various distributions \cite{Brody,CGR,Izzy,LH,BR,PR,Bogo} have been proposed and used to interpolate between the Poisson and Wigner limits. 
The most widely used is the Brody distribution \cite{Brody}:
\begin{subequations}\label{Eqn:Brody} 
\begin{equation}\label{Brodyp1}
P_{B}(z;q) = \alpha (q+1) z^{q} \exp\left( -\alpha z^{q+1}\right), 
\end{equation}  
where 
\begin{equation}\label{Brodyp2}
\alpha = \left[ \Gamma \left( \frac{q+2}{q+1}\right) \right]^{q+1},
\end{equation}
\end{subequations}
and $q\in[0,1]$ is the ``level repulsion exponent'' or ``Brody parameter''. This reduces to the Poisson distribution [Eq.~(\ref{Eqn:Poisson})] when $q=0$ and the Wigner distribution [Eq.~(\ref{Eqn:Wigner})] when $q=1$.
The Brody distribution has a long history in nuclear physics \cite{Brody2,nukandchaos} and in quantum chaos \cite{Stockmann,RobnikReview,Reichl} and has more recently emerged as an important and useful tool for analyzing the spectral statistics of complex networks \cite{NETs1,NETs2,NETs3,NETs4}. Outside of these established areas of application, the Brody distribution has been employed in many other fields (see, for example, Refs.~\cite{fancy1,fancy2}) and continues to enjoy widespread use \cite{recapps1,recapps2,recapps3,recapps4,recapps5}. 

Despite its successful and widespread use, the Brody distribution has no theoretical foundation in the context of eigenvalue statistics (i.e., it is not derived from any physical or mathematical theory). Equation (\ref{Eqn:Brody}) is essentially a surmise, but because it is a surmise based on the well-established phenomenon of level repulsion, it is often described as being a ``phenomenological'' \cite{Guhr} or ``semi-empirical'' \cite{NETs3} distribution (``heuristic'' \cite{nukandchaos} is another common descriptor). While the Brody distribution falls within the scope of random matrix theory (RMT) \cite{Mehta}, its association with RMT has thus far been groundless given the conspicuous absence of random-matrix models having Brody spacing distributions. In fact, the $2\times2$ model of Nieminen and Muche \cite{JohnandLutz} is (to the author's knowledge) the only model that has been put forward as a candidate; unfortunately the spacing distribution for their model is not exactly the Brody distribution. Indeed, a long-standing criticism of the Brody distribution is that it does not derive from any known random-matrix model. The existence question \cite{JohnandLutz} is resolved in the current paper, which presents a class of $2\times2$ random-matrix models that has the Brody distribution as its exact eigenvalue spacing distribution. The question of uniqueness is also addressed. 

It is important to state at the outset that the random matrices defined in this paper do not have the properties of traditional GOE or (more generally) Wigner matrices; in particular, the matrix elements are dependent and (in general) non-identically distributed. This is not a flaw since independent and identically-distributed (IID) matrix elements are model assumptions employed in classical RMT that can be dropped \cite{Anderson2008,Chak2015,Ben2015,AEK2017} without necessarily altering the classical RMT results obtained under IID conditions \cite{Chatter2006,Credner2008,EYY2012,Gotze2015,Naumov2015,Adamsky2016,Che2017,AEK2019}. This ``spectral universality'' under different non-IID conditions in theory applies only to large matrices, but as will be later shown using explicit examples, the elements of $2\times2$ random matrices need not be IID to uphold result (\ref{Eqn:Wigner}). To the author's knowledge, a general and rigorous proof of this specific fact does not exist in the literature. One other important feature of the random matrices considered here is that they are generally non-symmetric (when the matrices are real) and non-Hermitian (when the matrices are complex). GOE matrices owe their real and symmetric structure to assumptions and conditions that originate from traditional (i.e., Hermitian) quantum mechanics. In contrast (and for generality), the random-matrix model defined in this paper has no built-in quantum-mechanical assumptions or conditions.

From a statistics perspective, it has been recognized that the Brody distribution is related to the standard two-parameter Weibull distribution \cite{cW1,cW2,cW3} (the authors of Ref.~\cite{cW1} more specifically identified the former as a particular case of the latter). More precisely, the Brody distribution is a rescaled one-parameter Weibull distribution with unit mean \cite{cWs1,cWs2,cWs3}. The relevance of this relationship is that it offers a clue as to how to construct low-dimensional random-matrix models that will have Brody spacing distributions. For example, in the case of $2\times2$ matrices, if matrix elements can be assigned such that the discriminant of the characteristic polynomial of the random matrix is proportional to the square of a Weibull random variable, then the eigenvalue spacing distribution will be exactly the Brody distribution. The preceding can actually be achieved using several different constructions. One possible construction is to use (finite) sums of an exponential random variable raised to different $q$-dependent powers. A model based on this construction will be introduced and studied in this paper. 

The contents of the paper are as follows. In Section \ref{CRMMdef}, a general class of $2\times2$ correlated random-matrix models is defined. The eigenvalue spacing distribution for this class of models is then derived in Section \ref{NNSD} with the result being the Brody distribution. Terminology and notation for sub-classes are set out in Section \ref{SubModsDefns}. Three simple non-symmetric sub-classes of models with varying matrix-element structures are then defined and examined in Section \ref{NonSymMods}; the significance of each sub-class is discussed therein and illustrative examples are provided. Some simple symmetric sub-classes of models are presented in Section \ref{RSegsSCs}. Additional examples involving complex matrices and eigenvalues are given in Section \ref{cmplxySEC}. Some generalizations and extensions of the model introduced in Section \ref{CRMMdef} are considered in Section \ref{GensExts}. Numerical results for selected real and complex examples from Sections \ref{NonSymMods} and \ref{cmplxySEC} are then presented in Section \ref{NumEgs}. Comments (Section \ref{COMMs}), the Conclusion (Section \ref{CONCLSN}), and several Appendices (\ref{StatsStuff}-\ref{RayCalc}) complete the paper. 

\section{A General Class of $2\times2$ Correlated Multivariate Models}\label{CRMMdef}

Let $q\in[0,1]$ be a free parameter (which shall later be identified as the usual Brody parameter). Consider the general class of $2\times2$ random matrices of the form:
\begin{eqnarray}\label{MBG1}
   \mathcal{M}_n=\mathcal{A}\left ( \begin{array}{cc}
             \displaystyle g_1(X)+\sum_{m=1}^{n+1}k_mY^{p_{(m-1)}\over{q+1}} & ~~\left(\displaystyle\sum_{m=n+2}^{2n+2}k_mY^{p_{[m-(n+2)]}\over{q+1}}\right)h(T) \\ [0.6cm]
             \displaystyle{1 \over h(T)} \left(\sum_{m=2n+3}^{3n+3}k_mY^{p_{[m-(2n+3)]}\over{q+1}}\right) & ~~~\displaystyle g_2(X)+\sum_{m=3n+4}^{4n+4}k_mY^{p_{[m-(3n+4)]}\over{q+1}}
           \end{array} \right),
\end{eqnarray}
where $m$ and $n$ are positive integers, $\{T,X,Y\}$ are independent random variables whose distributions will be specified below, $\{g_1(X),g_2(X)\}$ are functions of $X$ whose difference is a fixed (real or complex) number for all supported values of $X$ (i.e., $g_1(x)-g_2(x)=\eta\in\mathbb{C}$, $\forall x\in\text{supp}(X)$), $h(T)$ is any function of $T$ having the property $h(T)\neq0$ for all supported values of $T$, $\{k_m:m=1,2,3,\ldots,4n+4\}$ are a constrained set of (real or complex) constants, and $\left\{{p_1},{p_2},\ldots,{p_n}\right\}$ ($p_0=0$) are a constrained set of unequal real constants that can (but need not) have parametric dependencies. The prefactor $\mathcal{A}$ (outside the matrix) is an overall multiplicative factor for each of the matrix elements that can (in general) depend on any of the prior-defined constants or variables. The distributions of $\{T,X,Y\}$ are as follows. Let $Y\sim\text{Exp}(\sigma_{\text{e}})$, that is, let $Y$ have an exponential distribution with scale parameter $\sigma_\text{e}>0$:\footnote{In this paper, the common notation $f_V(v)$ will be used to refer to the p.d.f.~of an arbitrary random variable $V$ with lower-case $v$ then denoting the density variable corresponding to $V$.} 
\begin{equation}\label{rayden}
f_Y(y;\sigma_\text{e})={1\over\sigma_\text{e}}\exp\left(-{y\over{\sigma_\text{e}}}\right),~y>0.
\end{equation}
In the vernacular of RMT, $Y$ has a ``Poisson distribution'' (in the density variable $y$) when $\sigma_\text{e}=1$. The random variables $X$ and $T$ can have any arbitrary continuous distributions $f_X(x)$ and $f_T(t)$, respectively (including exponential distributions). Thus, apart from the functions $\{g_1(X),g_2(X),h(T)\}$ (and a possibly non-constant prefactor $\mathcal{A}$), the matrix elements are constrained sums of power-transformed exponential random variables. As such, the matrix elements are generally not independent nor identically distributed. Lastly, the discriminant of the characteristic polynomial of $\mathcal{M}_n$, denoted by $\mathcal{D}(\mathcal{M}_n)$, must satisfy the condition:
\begin{equation}\label{condhs}
\mathcal{D}(\mathcal{M}_n)\equiv\left[\text{Tr}(\mathcal{M}_n)\right]^2-4\det(\mathcal{M}_n)=kY^{2\over{q+1}},
\end{equation}
where the ``discriminant constant'' $k$ is a positive real constant.\footnote{The constant $k$, which will depend on the constants $\{k_m\}$ in model (\ref{MBG1}), could also be negative, in which case, the eigenvalues are not real; this case will be considered in Section \ref{CCevsSec}.} As will become clear in Section \ref{NNSD}, condition (\ref{condhs}) is a sufficient condition for the eigenvalue spacing distribution of $\mathcal{M}_n$ to be the Brody distribution.  

\section{Derivation of the Eigenvalue Spacing Distribution}\label{NNSD}

Let $\lambda_-$ and $\lambda_+$ denote the two eigenvalues of $\mathcal{M}_n$. These two eigenvalues can be obtained from the general formula
\begin{equation}\label{EVs}
\lambda_\pm={\text{Tr}(\mathcal{M}_n)\pm\sqrt{\mathcal{D}(\mathcal{M}_n)}\over2},
\end{equation}
where the discriminant $\mathcal{D}(\mathcal{M}_n)$ is as defined in Eq.~(\ref{condhs}). 
The spacing between the two eigenvalues, denoted here by the random variable $S$, is subsequently given by:
\begin{equation}\label{spacsG}
S\equiv\lambda_+-\lambda_-=\sqrt{\mathcal{D}(\mathcal{M}_n)}.
\end{equation}
Note that Eq.~(\ref{spacsG}) is valid only when the eigenvalues are real, which will be the case for almost all sub-classes of models considered in this paper (exceptions occur in Section \ref{CCevsSec}). 

Utilizing the fact that the support of $Y$ is $(0,\infty)$ (note the restriction on the density variable $y$ in Eq.~(\ref{rayden})), application of Eq.~(\ref{spacsG}) to the model defined by Eqs.~(\ref{MBG1})-(\ref{condhs}) yields:
\begin{equation}\label{spacs}
S=\sqrt{k}Y^{1\over{q+1}},
\end{equation}
where $k>0$ is the discriminant constant. 
Note that $S$ is independent of the functions $\{g_1(X),g_2(X),h(T)\}$. As alluded to in Section \ref{CRMMdef}, the simple functional form for $S$ given by Eq.~(\ref{spacs}) is due to condition (\ref{condhs}).

To determine the distribution of $S$, it is mathematically efficient to exploit the well-known functional relation between exponential and Weibull random variables (see, for example, Ref.~\cite{Norman}): if $Y$ has an exponential distribution with scale parameter $\sigma_\text{e}>0$, then the random variable $W\equiv Y^{1/\tau}$, where $\tau>0$, has a Weibull distribution with scale parameter $\sigma_\text{e}$ and shape parameter $\tau$. Thus, 
\begin{equation}\label{RayisGG}
W\equiv Y^{1\over{q+1}}\sim\text{Weibull}\left(\kappa=\sigma_\text{e},\tau=q+1\right),
\end{equation}
where the general notation for the scale and shape parameters of the Weibull distribution given in Appendix \ref{AppWBL} has been utilized. 
Using (\ref{GGD}) with parameter $\tau=q+1$, the p.d.f.~of $S$ is thus:
\begin{eqnarray}\label{preBrody}
f_S(s)&=&f_{\sqrt{k}W}(s)={1\over \sqrt{k}}f_W\left(w={s\over \sqrt{k}};\kappa=\sigma_\text{e},\tau=q+1\right) \nonumber \\
&=&{{(q+1)}\over{\sqrt{k}\sigma_\text{e}}}\left(s\over\sqrt{k}\right)^q\exp\left[-{1\over\sigma_\text{e}}\left(s\over\sqrt{k}\right)^{q+1}\right],
\end{eqnarray}
where $\sigma_\text{e}$ is the exponential scale parameter in Eq.~(\ref{rayden}), and $k$ is the discriminant constant. Equation (\ref{preBrody}) is the distribution of the eigenvalue spacings of $\mathcal{M}_n$. This is not quite the Brody distribution because the spacing $S$ has not yet been rescaled by its mean $\mu_S$. Recall that we ultimately seek the distribution of the standardized or mean-scaled spacing
\begin{equation}\label{soversbar}
Z\equiv {S/\mu_S}. 
\end{equation} 
From (\ref{meanW}), it follows that the mean of $f_S(s)$ is:
\begin{equation}\label{meanS}
\mu_S=\mu_{\sqrt{k}W}=\sqrt{k}\mu_W(\kappa=\sigma_\text{e},\tau=q+1)=\sqrt{k}\sigma_\text{e}^{1\over{q+1}}\Gamma\left({{q+2}\over{q+1}}\right). \quad
\end{equation} 
The p.d.f.~of $Z$ can then be obtained from a simple change of variables as follows:
\begin{equation}\label{toBrody}
f_Z(z)=f_S(s=\mu_Sz)\times\mu_S.
\end{equation}
Performing the necessary algebra then yields:
\begin{equation}\label{atBrody}
f_Z(z)=P_B(z;q),
\end{equation}
which is the Brody distribution [Eq.~(\ref{Eqn:Brody})]. The parameter $q$ in model (\ref{MBG1}) (and all sub-classes of models that derive from model (\ref{MBG1})) is therefore the usual Brody parameter. It is important to note that the unscaled spacing distribution (\ref{preBrody}) depends on both the exponential scale parameter $\sigma_\text{e}$ in Eq.~(\ref{rayden}) and the discriminant constant $k$, whereas the mean-scaled spacing distribution (\ref{atBrody}) has no such dependencies. 

\section{Sub-Classes of $\mathcal{M}_n$: Terminology and Notation}\label{SubModsDefns}

Many different sub-classes of models can be constructed from the general multivariate model defined in Section \ref{CRMMdef}. For future reference, the term ``cropped'' shall be used to refer to models having all of the following properties: (i) $g_1(X)=g_2(X)=0$, (ii) $\mathcal{A}=h(T)=1$, and (iii) no lone constants (in the summations); models having only properties (i) and (ii) will be referred to as ``pruned''; models having properties (ii) and (iii) and (instead of (i)) the property $g_1(X)=g_2(X)\neq0$ will be referred to as ``trimmed''. Low-order sub-classes of models (e.g., $n=2,~3$, or $4$) offer a degree of parametric versatility (via the existence of one independent parameter) and are algebraically simple. Higher-order models (e.g., $n>4$) can be more parametrically versatile but they are also more algebraically complicated. For example, imposing condition (\ref{condhs}) on a cropped trivariate model
\begin{eqnarray}\label{MBGeg3t}
    \mathcal{M}^{(0)}_3=\left ( \begin{array}{cc}
             \displaystyle\sum_{m=1}^3k_mY^{p^{}_{m_{}}\over{q+1}} & ~~\displaystyle\sum_{m=4}^6k_mY^{p_{(m-3)}\over{q+1}} \\ [0.5cm]
             \displaystyle\sum_{m=7}^9k_mY^{p_{(m-6)}\over{q+1}} & ~~\displaystyle\sum_{m=10}^{12}k_mY^{p_{(m-9)}\over{q+1}}
           \end{array} \right)
\end{eqnarray}
generates an underdetermined quadratic polynomial system consisting of six quadratic equations in 12 unknowns (i.e., the constants $\{k_m:m=1,\ldots,12\}$), whereas imposing condition (\ref{condhs}) on a cropped quadvariate model
\begin{eqnarray}\label{MBGeg4t}
   \mathcal{M}^{(0)}_4=\left ( \begin{array}{cc}
             \displaystyle\sum_{m=1}^4k_mY^{p^{}_{m_{}}\over{q+1}} & ~~\displaystyle\sum_{m=5}^8k_mY^{p_{(m-4)}\over{q+1}} \\ [0.5cm]
             \displaystyle\sum_{m=9}^{12}k_mY^{p_{(m-8)}\over{q+1}} & ~~\displaystyle\sum_{m=13}^{16}k_mY^{p_{(m-12)}\over{q+1}}
           \end{array} \right)
\end{eqnarray} 
generates an underdetermined quadratic polynomial system consisting of ten quadratic equations in 16 unknowns (i.e., the constants $\{k_m:m=1,\ldots,16\}$). 

Low-order models are sufficiently versatile and algebraically simple and will hence be the focus of this paper. In the following sections, several different sub-classes of models will be studied. As a notational convenience, the constrained constants $\left\{{p_1},{p_2},{p_3},\ldots\right\}$ appearing in the general model (\ref{MBG1}) will be denoted by $\left\{a,b,c,\ldots\right\}$ and the constants $\{k_m\}$ appearing in model (\ref{MBG1}) will be replaced by a simpler set of constants $\{c_j\}$. In this way, the sub-classes can be written as stand-alone models without explicit reference to the general model (\ref{MBG1}). To illustrate this notational scheme, consider the sub-class of   
\begin{eqnarray}\label{MBGegTtoA1}
   \mathcal{M}_2=\mathcal{A}\left ( \begin{array}{cc}
             \displaystyle g_1(X)+\sum_{m=1}^{3}k_mY^{p_{(m-1)}\over{q+1}} & ~~\displaystyle\left(\sum_{m=4}^{6}k_mY^{p_{(m-4)}\over{q+1}}\right)h(T) \\ [0.5cm]
             \displaystyle {1 \over h(T)} \left(\sum_{m=7}^{9}k_mY^{p_{(m-7)}\over{q+1}}\right) & ~~\displaystyle g_2(X)+\sum_{m=10}^{12}k_mY^{p_{(m-10)}\over{q+1}}
           \end{array} \right)
\end{eqnarray} 
where the constants $\{k_2,k_3,k_6,k_{8},k_{11},k_{12}\}$ are all identically zero and $\mathcal{A}=h(T)=1$ (referred to here as sub-class E of class $\mathcal{M}_2$):
\begin{subequations}\label{MBGegTtoA}
\begin{eqnarray}\label{MBGegTtoA2}
   \mathcal{M}_2^{(E)}=\left ( \begin{array}{cc}
             g_1(X)+k_1 & ~k_4+k_5Y^{p_1\over{q+1}} \\
             k_7+k_{9}Y^{p_2\over{q+1}} & ~g_2(X)+k_{10}
           \end{array} \right).
\end{eqnarray} 
Notwithstanding notation, sub-class E is a viable class of models in its own right and need not be regarded as a sub-class of class $\mathcal{M}_2$. Conceptualizing it as such, sub-class E can be alternatively written (without notational reference to $\mathcal{M}_2$) in the general form:
\begin{eqnarray}\label{MBGegTtoA3}
   \mathcal{M}_E=\left ( \begin{array}{cc}
             c_1+g_1(X) & ~c_2+c_3Y^{a\over{q+1}} \\
             c_4+c_5Y^{b\over{q+1}} & c_6+g_2(X)
           \end{array} \right),
\end{eqnarray}
\end{subequations}
where $a \neq b$, and the constants $\{c_j:j=1,\ldots,6\}$ are generally different from zero. For the eigenvalues of $\mathcal{M}_E$ to have a Brody spacing distribution, various conditions on the constants $\{c_j\}$ and the power constants $\{a,b\}$ can be imposed such that condition (\ref{condhs}) is satisfied. For example, the random matrix
\begin{eqnarray}\label{MBsim2}
   \mathcal{M}_E^{(1)}=\left ( \begin{array}{cc}
             \tan^{-1}(X)+q & ~c+Y^{-{2\over{q+1}}} \\
             -\left(q+{\pi\over4}\right)^2Y^{2\over{q+1}} & ~-\cot^{-1}(X)-q
           \end{array} \right),~c<0 
\end{eqnarray}
is one member of sub-class E. Note that in the above example, the off-diagonal matrix elements are independent of the diagonal matrix elements. 

\section{Simple Non-Symmetric Models}\label{NonSymMods}

Simple trimmed, cropped, and pruned sub-classes of models are (under condition (\ref{condhs})) typically non-symmetric. The following three subsections consider (in turn) a low-order example of each aforementioned type. The understanding is that these non-symmetric models are among the most basic $2\times2$ (correlated) models possessing Brody spacing distributions and that (if desired) the random functions $\{g_1(X),g_2(X),h(T)\}$ can be judiciously added to incorporate additional randomization (i.e., reduce correlation) and/or impose independence between particular matrix elements. 

\subsection{Sub-Class A}\label{subbyA}

Sub-Class A consists of random matrices of the (trimmed) form: 
\begin{subequations}\label{MB}
\begin{eqnarray}\label{MB1}
   \mathcal{M}_A=\left ( \begin{array}{cc}
             X+c_1Y^{1\over{q+1}} & c_2Y^{a\over{q+1}} \\
             c_3Y^{b\over{q+1}} & X+c_4Y^{1\over{q+1}}
           \end{array} \right),
\end{eqnarray}
where $\{c_1,c_2,c_3,c_4\}$ are arbitrary real constants subject to the condition 
\begin{eqnarray}\label{MB2}
(c_1-c_4)^2+4c_2c_3>0,
\end{eqnarray}
which ensures that the eigenvalues are real and non-degenerate\footnote{A separate condition on the constants could be added to ensure that all eigenvalues are positive, but this is not necessary.},  and the unequal constants $\{a,b\}$ are subject to the condition
\begin{eqnarray}\label{MB3}
a+b=2 \quad \text{if}~c_2\neq0~\text{and}~c_3\neq0, 
\end{eqnarray}
\end{subequations}
but are otherwise arbitrary. 

The diagonal matrix elements are generally dependent for arbitrarily prescribed non-zero values of $c_1$ and $c_4$; they are trivially independent when $X=0$ and one of $\{c_1,c_4\}$ is also zero. Similarly, the off-diagonal matrix elements are trivially independent when one of $\{c_2,c_3\}$ is zero or when one of $\{a,b\}$ is zero. For arbitrary non-zero values of $\{c_2,c_3\}$, the diagonal and off-diagonal matrix elements are non-trivially independent of each other only when both $c_1$ and $c_4$ are zero; the upper (lower) off-diagonal matrix element is trivially independent of the diagonal matrix elements when $a=0$ ($b=0$). Hence, under general conditions, any pair of matrix elements is mutually dependent. The matrix elements are also generally non-identically distributed; the only exception occurs when $c_1=c_4$ in which case the diagonal elements are identically distributed. 

\subsubsection{Illustrative Examples}

Under the assumption that $\{c_2,c_3\}$ are both non-zero, choosing
\begin{subequations}\label{MBegs} 
\begin{eqnarray}\label{QsnsEG}
a(q)=-q,~b(q)=q+2 
\end{eqnarray}
yields the following sub-class of random matrices:
\begin{eqnarray}\label{MBnsEG}
   \mathcal{M}^{(1)}_{A}=\left ( \begin{array}{cc}
             X+c_1Y^{1\over{q+1}} & c_2Y^{-{{q}\over{q+1}}} \\
             c_3Y^{{{q+2}\over{q+1}}} & X+c_4Y^{1\over{q+1}}
           \end{array} \right).
\end{eqnarray}
\end{subequations}

A cropped subset of sub-class A with vanishing trace can be obtained by setting $X=0$ and fixing $c_4=-c_1$ ($c_1$ is then a free constant). For example, choosing
\begin{subequations}\label{MBegsTr0} 
\begin{eqnarray}\label{QsnsTr0EG}
a(q)=\sec^2(q)+\csc^2(q+1),~b(q)=-\left(\tan^2(q)+\cot^2(q+1)\right) 
\end{eqnarray}
yields the following (cropped) zero-trace sub-class of random matrices:
\begin{eqnarray}\label{MBnsTr0EG}
   \mathcal{M}^{(2)}_{A}=\left ( \begin{array}{cc}
             c_1Y^{1\over{q+1}} & ~c_2Y^{\sec^2(q)+\csc^2(q+1)\over{q+1}} \\
             c_3Y^{-{\tan^2(q)+\cot^2(q+1)\over{q+1}}} & -c_1Y^{1\over{q+1}}
           \end{array} \right),
\end{eqnarray}
\end{subequations}
where the constants $\{c_1,c_2,c_3\}$ must satisfy the condition $c^2_1+c_2c_3>0$. 

\subsubsection{Significance of Sub-Class A}

For any given order $n\ge2$, there exist sub-classes of trimmed models whose off-diagonal elements are composed of distinct random variables (i.e., the off-diagonal elements are not different linear combinations of the same random variable(s)). The significance of sub-class A (a sub-class of class $\mathcal{M}_3$) is that it is (essentially) the simplest such sub-class.\footnote{The parentheses are to account for the fact that the simplest such sub-class, strictly speaking, is the one obtained by choosing $c_1=c_4=0$ in model (\ref{MB}).}    

For comparison, a slightly more complicated sub-class of trimmed models whose off-diagonal elements are composed of distinct random variables is the sub-class
\begin{subequations}\label{MBcomp}
\begin{eqnarray}\label{MBcomp1}
   \mathcal{M}_I=\left ( \begin{array}{cc}
             X+c_1Y^{c\over{q+1}} & ~c_2Y^{a\over{q+1}}+c_5Y^{c\over{q+1}} \\
             c_3Y^{b\over{q+1}} & X+c_4Y^{c\over{q+1}}
           \end{array} \right),
\end{eqnarray}
where $\{c_j:j=1,\ldots,5\}$ are arbitrary real constants subject to the conditions 
\begin{eqnarray}\label{MBcomp2}
(c_1-c_4)^2+4c_2c_3=0,~c_3c_5>0,
\end{eqnarray}
and the constants $\{a,b,c\}$ are unequal and subject to the conditions
\begin{eqnarray}\label{MBcomp3}
a+b=2c,~b+c=2,~c\neq0, 
\end{eqnarray}
\end{subequations}
but are otherwise arbitrary. $\mathcal{M}_I$ reduces to $\mathcal{M}_A$ when $c_5=0$ and $c=1$ (the second condition in (\ref{MBcomp3}) must also be dropped). As a simple example, one member of sub-class I is the random matrix:
\begin{eqnarray}\label{MBcomp1EG1}
   \mathcal{M}^{(1)}_I=\left ( \begin{array}{cc}
             X-Y^{3\over{q+1}} & ~-Y^{7\over{q+1}}+Y^{3\over{q+1}} \\
             Y^{-{1\over{q+1}}} & X+Y^{3\over{q+1}}
           \end{array} \right). 
\end{eqnarray}
The power constants $\{a,b,c\}$ could as well be prescribed to be $q$-dependent. For example,
\begin{eqnarray}\label{MBcomp1EG2}
   \mathcal{M}^{(2)}_I=\left ( \begin{array}{cc}
             X+Y^{\sin^2\left(\phi\right)\over{q+1}} & ~Y^{\sin^2\left(\phi\right)-2\cos^2\left(\phi\right)\over{q+1}}-Y^{\sin^2\left(\phi\right)\over{q+1}} \\ [0.1cm]
             -Y^{{\sin^2\left(\phi\right)+2\cos^2\left(\phi\right)\over{q+1}}} & X-Y^{\sin^2\left(\phi\right)\over{q+1}}
           \end{array} \right), 
\end{eqnarray}
where $\phi\equiv\phi(q)$ can be any real function of $q$ that is continuous and non-zero for all $q\in[0,1]$ (e.g., $\phi(q)=\pi^{q-1}$). 

\subsubsection{Generalization of a Special Case}

When either $c_2$ or $c_3$ is zero, sub-class A matrices can be generalized as follows:
\begin{subequations}\label{MB1c23eq0}
\begin{eqnarray}\label{MB1c20}
   \mathcal{M}^{\text{(g)}}_A(c_2=0)=\left ( \begin{array}{cc}
             X+c_1Y^{1\over{q+1}} & 0 \\
             V & X+c_4Y^{1\over{q+1}}
           \end{array} \right),
\end{eqnarray}
\begin{eqnarray}\label{MB1c30}
   \mathcal{M}^{\text{(g)}}_A(c_3=0)=\left ( \begin{array}{cc}
             X+c_1Y^{1\over{q+1}} & V \\
             0 & X+c_4Y^{1\over{q+1}}
           \end{array} \right),
\end{eqnarray}
\end{subequations}
where the random variable $V$ is independent of both $X$ and $Y$ and can have any continuous distribution $f_V(v)$. It should be noted that when one or both of $\{c_2,c_3\}$ are zero, the result is a tridiagonal matrix, in which case, the off-diagonal matrix elements have no effect on the eigenvalue spacing distribution (the latter fact can also be deduced from direct application of Eq.~(\ref{spacsG})). It should also be noted that while the diagonal matrix elements are generally dependent all other pairs of matrix elements are mutually independent. 

\subsubsection{A Related Class of Models}

It is worth mentioning the existence of a similar class of models that does not derive from model (\ref{MBG1}), namely, the class of random matrices:
\begin{eqnarray}\label{UnrelClass}
   \mathcal{M}_J=\left ( \begin{array}{cc}
             c_1d_1(X)Y^{1\over{q+1}} & ~c_2Y^{a\over{q+1}} \\
             c_3Y^{b\over{q+1}} & ~c_1d_2(X)Y^{1\over{q+1}}
           \end{array} \right),
\end{eqnarray}
where the diagonal functions $\{d_1(X),d_2(X)\}$ are any two real functions of the random variable $X$ that satisfy $d_1(X)-d_2(X)=\eta$, where $\eta$ is a real-valued constant, the unequal constants $\{a,b\}$ satisfy condition (\ref{MB3}) (but are otherwise arbitrary), and the real constants $\{c_1,c_2,c_3\}$ satisfy the condition $c^2_1\eta^2+4c_2c_3>0$. Members of this class also satisfy condition (\ref{condhs}) and hence have Brody spacing distributions.

An example is the random matrix:
\begin{eqnarray}\label{UnrelClassEG}
   \mathcal{M}^{(1)}_J=\left ( \begin{array}{cc}
             \cos^2(X)Y^{1\over{q+1}} & ~Y^{4\sin^{-1}(q-1)\over{\pi(q+1)}} \\
             Y^{4\cos^{-1}(q-1)\over{\pi(q+1)}} & ~{1\over2}\cos(2X)Y^{1\over{q+1}}
           \end{array} \right).
\end{eqnarray}

\subsection{Sub-Class B}

Sub-Class B consists of random matrices of the (cropped) form: 
\begin{eqnarray}\label{MBmBp1}
   \mathcal{M}_B=\left ( \begin{array}{cc}
             c_1Y^{a\over{q+1}} & c_2Y^{a\over{q+1}}+c_3Y^{b\over{q+1}} \\
             c_4Y^{a\over{q+1}}+c_5Y^{b\over{q+1}} & c_6Y^{b\over{q+1}}
           \end{array} \right),
\end{eqnarray}
where $a \neq b$, and the constants $\{c_j:j=1,\ldots,6\}$ are assumed to be real and such that $\{c_1,c_6\}$ are both non-zero. The discriminant of the characteristic polynomial of $\mathcal{M}_B$ is:
\begin{subequations}\label{discrimMB}
\begin{equation}
\mathcal{D}({\mathcal{M}_B})=\mathcal{C}_1Y^{2a\over{q+1}}+\mathcal{C}_2Y^{a+b\over{q+1}}+\mathcal{C}_3Y^{2b\over{q+1}},
\end{equation}
where
\begin{eqnarray}\label{MBmBp2}
&&~~\mathcal{C}_1 \equiv c^2_1+4c_2c_4, \label{MBmBp21} \\
&&~~\mathcal{C}_2 \equiv -2c_1c_6+4(c_2c_5+c_3c_4), \label{MBmBp22} \\
&&~~\mathcal{C}_3 \equiv c^2_6+4c_3c_5. \label{MBmBp23} 
\end{eqnarray}
\end{subequations}
The constants $\{a,b\}$ are thus subject to the following conditions based on the values of the above-defined constants (but are otherwise arbitrary): 
\begin{subequations}\label{MBtunies}
\begin{eqnarray}
&a=1&~~~\text{if}~\{\mathcal{C}_1>0,\mathcal{C}_2=0,\mathcal{C}_3=0\}, \label{MBtunies1} \\
&a+b=2&~~~\text{if}~\{\mathcal{C}_1=0,\mathcal{C}_2>0,\mathcal{C}_3=0\}, \quad \quad \quad \label{MBtunies2} \\
&b=1&~~~\text{if}~\{\mathcal{C}_1=0,\mathcal{C}_2=0,\mathcal{C}_3>0\}. \label{MBtunies3} 
\end{eqnarray}
\end{subequations}
As defined, sub-class B does not allow for real symmetric matrices since case (\ref{MBtunies1}) demands that constants $\{c_3,c_5\}$ be of opposite sign and case (\ref{MBtunies3}) demands that constants $\{c_2,c_4\}$ be of opposite sign; case (\ref{MBtunies2}) demands both the former and the latter.\footnote{These statements presume that none of the constants in question are zero.} Thus, under general conditions (e.g., none of the constants are zero), the matrix elements are dependent and non-identically distributed. For future reference, cases (\ref{MBtunies1})-(\ref{MBtunies3}) will be referred to as Cases I-III, respectively. 
When one of $\{c_3,c_5\}$ is zero (i.e., no $Y^{b\over{q+1}}$ term in one of the off-diagonal elements), the result is a simpler sub-class of matrices subsumed by Case III. Similarly, when one of $\{c_2,c_4\}$ is zero (i.e., no $Y^{a\over{q+1}}$ term in one of the off-diagonal elements), the result is a simpler sub-class of matrices subsumed by Case I.

Conditions (\ref{MBtunies}) may seem restrictive, but they actually constitute an underdetermined quadratic polynomial system (six variables, two quadratic equations, and one quadratic inequality constraint), and such systems are known to be either inconsistent (i.e., have no solution) or possess an infinite number of real and complex solutions \cite{Commy}. In the present case, one can easily find a few families of solutions using simple trial and error, and hence the system is not inconsistent (i.e., there are an infinite number of solutions). It is worth noting that the effect of the quadratic inequality is only to impose some restrictions on the (infinity of) solutions to the two quadratic equations. The full set of solutions to this quadratic system can be determined (with the aid of computer algebra software) and expressed in parametric form but it is not necessary since the only mathematical requirement (for sub-class B to be well defined) is that the polynomial system not be inconsistent (i.e., the conditions on the constants can actually be satisfied). Some illustrative examples are given below. 

\subsubsection{Illustrative Examples}

If the power constants $\{a,b\}$ are assumed to be $q$-dependent, which is the simplest parametrization of sub-class B, then examples for Cases I, III, and II (respectively) are:
\begin{subequations}\label{MBBegs}
\begin{eqnarray}\label{MBBeg1}
   \mathcal{M}^{(1)}_{B(I)}=\left ( \begin{array}{cc}
             Y^{1\over{q+1}} & 2Y^{1\over{q+1}}-{1\over2}Y^{u(q)} \\
             Y^{1\over{q+1}}+{1\over2}Y^{u(q)} & Y^{u(q)}
           \end{array} \right),
\end{eqnarray}
where the exponent $u(q)$ can be any real function of $q$ that is continuous for all $q\in[0,1]$,
\begin{eqnarray}\label{MBBeg3}
   \mathcal{M}^{(1)}_{B(III)}=\left ( \begin{array}{cc}
             2Y^{v(q)} & Y^{v(q)}+2Y^{1\over{q+1}} \\
             -Y^{v(q)}+5Y^{1\over{q+1}} & 3Y^{1\over{q+1}}
           \end{array} \right),
\end{eqnarray}
where the exponent $v(q)$ can be any real function of $q$ that is continuous for all $q\in[0,1]$, and 
\begin{eqnarray}\label{MBBeg2}
   \mathcal{M}^{(1)}_{B(II)}=\left ( \begin{array}{cc}
             -2Y^{a(q)\over{q+1}} & ~{-{3\over2}}Y^{a(q)\over{q+1}}+{3\over2}Y^{b(q)\over{q+1}} \\ [0.1cm]
             {2\over3}Y^{a(q)\over{q+1}}-{3\over2}Y^{b(q)\over{q+1}} & -3Y^{b(q)\over{q+1}}
           \end{array} \right),
\end{eqnarray}
where $\{a(q),b(q)\}$ can be any real unequal functions of $q$ that satisfy condition (\ref{MBtunies2}) and are continuous on the interval $[0,1]$. 
As the above examples illustrate, there is a rather large degree of flexibility in assigning the $q$ dependencies of the exponents of $Y$.
\end{subequations}

\subsubsection{Significance of Sub-Class B}\label{SignifB}

For any given order $n\ge2$, there exist sub-classes of cropped models whose diagonal elements are composed of strictly distinct random variables. The significance of sub-class B (a sub-class of class $\mathcal{M}_2$) is that it is the simplest such sub-class. Outside of cropped models, simpler sub-classes having the aforementioned matrix-element structure certainly exist, for example, pruned models of the forms: 
\begin{subequations}\label{Bsig1}
\begin{eqnarray}\label{MBsig1}
   \left ( \begin{array}{cc}
             c_1Y^{a\over{q+1}}+c_2 & ~c_3Y^{a \vee b \vee c\over{q+1}}+c_4 \\
             c_5Y^{c\over{q+1}} +c_6 & c_7Y^{b\over{q+1}}+c_8
           \end{array} \right),
\end{eqnarray}
where 
\begin{equation}\label{condysig1}
c_1\neq0,c_7\neq0,a\neq0,b\neq0,a \neq b,c\neq0, 
\end{equation}
\end{subequations}  
and the real constants $\{c_j:j=1,\ldots,8\}$ and power constants $\{a,b,c\}$ are subject to various sets of conditions (beyond those given in (\ref{condysig1})) such that condition (\ref{condhs}) is satisfied. Note that $a \vee b \vee c$ appearing in the upper off-diagonal element denotes that either $a$, $b$, or $c$ can be chosen as the numerator of the exponent (three distinct sub-classes of models are implied by (\ref{Bsig1})). Although simpler in form, pruned models (\ref{Bsig1}) are less versatile than sub-class B since only certain discrete values of $\{a,b,c\}$ in models (\ref{Bsig1}) can satisfy condition (\ref{condhs}). For example, the model obtained by choosing $c$ as the numerator of the exponent in the upper off-diagonal element of (\ref{MBsig1}), that is, the pruned model 
\begin{eqnarray}\label{Bsig2}
   \left ( \begin{array}{cc}
             c_1Y^{a\over{q+1}}+c_2 & c_3Y^{c\over{q+1}}+c_4 \\
             c_5Y^{c\over{q+1}} +c_6 & c_7Y^{b\over{q+1}}+c_8
           \end{array} \right) 
\end{eqnarray} 
can (subject to stipulations (\ref{condysig1})) satisfy condition (\ref{condhs}) only when $\{a=1/3,b=1,c=2/3\}$ or when $\{a=1,b=1/3,c=2/3\}$. In contrast, at least one the power constants $\{a,b\}$ in sub-class B can assume a continuous set of real values. 

It can be shown and is worth noting that pruned models of the forms
\begin{eqnarray}\label{MBsig2}
   \left ( \begin{array}{cc}
             c_1Y^{a\over{q+1}}+c_2 & c_3Y^{a \vee b\over{q+1}}+c_4 \\
             c_5Y^{a \vee b\over{q+1}} +c_6 & c_7Y^{b\over{q+1}}+c_8
           \end{array} \right)
\end{eqnarray}
satisfying condition (\ref{condhs}) do not exist (assuming that no off-diagonal element is identically zero and stipulations (\ref{condysig1}) are respected). As before, the $a \vee b$ appearing in the off-diagonal elements of models (\ref{MBsig2}) denotes that either $a$ or $b$ can be chosen as the numerator of the pertinent exponent (four distinct sub-classes of models are implied by (\ref{MBsig2})). Care should be taken to observe that the special members of (\ref{MBsig2}) for which $c_2=c_4=c_6=c_8=0$ do not belong to sub-class B. 

\subsubsection{Generalizations of Cases I and III}\label{GenCIandIIIModB}

Cases I and III of sub-class B can be generalized as follows. Let 
\begin{subequations}\label{MBGEN1}
\begin{eqnarray}\label{MBGEN1A}
   \mathcal{M}^{(\text{g})}_{B(I)}=\left ( \begin{array}{cc}
             c_1Y^{1\over{q+1}} & c_2Y^{1\over{q+1}}+c_3V \\
             c_4Y^{1\over{q+1}}+c_5V & c_6V
           \end{array} \right),
\end{eqnarray}
where the random variable $V$ is independent of $Y$ and can have any continuous distribution $f_V(v)$. Then 
\begin{equation}\label{MBBGENDs1}
\mathcal{D}\left(\mathcal{M}^{(\text{g})}_{B(I)}\right)\propto Y^{2\over{q+1}}~~~\text{if}~\{\mathcal{C}_1>0,\mathcal{C}_2=0,\mathcal{C}_3=0\}.  
\end{equation}
\end{subequations}
Similarly, let
\begin{subequations}\label{MBBGEN2}
\begin{eqnarray}\label{MBBGEN2A}
   \mathcal{M}^{(\text{g})}_{B(III)}=\left ( \begin{array}{cc}
             c_1V & c_2V+c_3Y^{1\over{q+1}} \\
             c_4V+c_5Y^{1\over{q+1}} & c_6Y^{1\over{q+1}}
           \end{array} \right).
\end{eqnarray} 
Then 
\begin{equation}\label{MBBGENDs3}
\mathcal{D}\left(\mathcal{M}^{(\text{g})}_{B(III)}\right)\propto Y^{2\over{q+1}}~~~\text{if}~\{\mathcal{C}_1=0,\mathcal{C}_2=0,\mathcal{C}_3>0\}. 
\end{equation}
\end{subequations}
Note that in each of models $\mathcal{M}^{(\text{g})}_{B(I)}$ and $\mathcal{M}^{(\text{g})}_{B(III)}$ the diagonal matrix elements are mutually independent. 

\subsection{Sub-Class C}

Sub-class C consists of random matrices of the (pruned) form: 
\begin{eqnarray}\label{MCcomp1}
   \mathcal{M}_C=\left ( \begin{array}{cc}
             c_1Y^{a\over{q+1}} & ~c_2+c_3Y^{b\over{q+1}} \\
             c_4+c_5Y^{c\over{q+1}} & c_6Y^{d\over{q+1}}
           \end{array} \right),
\end{eqnarray}
where the real constants $\{c_j:j=1,\ldots,6\}$ and power constants $\{a,b,c,d\}$ are subject to various sets of conditions such that condition (\ref{condhs}) is satisfied. The constants are also subject to the following restrictions: (i) $\left\{c_1,c_3,c_5,c_6\right\}$ are non-zero; and (ii) $\left\{a,b,c,d\right\}$ are distinct.\footnote{Note that, by virtue of their definitions, sub-classes A, B, and C, are disjoint.} 

\subsubsection{Illustrative Examples}\label{SubCegs}

If the real constants $\{c_j\}$ satisfy the conditions
\begin{subequations}\label{MCegGen}   
\begin{eqnarray}
c_1^2+4c_3c_5=0,~c_2=0, \label{MCegGen1a} \\
c_6^2+4c_3c_4=0,~c_1c_6<0, \label{MCegGen1b}
\end{eqnarray}
and the unequal power constants $\{a,b,c,d\}$ satisfy the conditions
\begin{eqnarray}\label{MCegGen2}
2a=b+c,~a+d=2,~b=2d,
\end{eqnarray}
\end{subequations}
then condition (\ref{condhs}) will be satisfied. An infinite number of real solutions exist for systems (\ref{MCegGen}). An example (satisfying conditions (\ref{MCegGen})) is the random matrix:
\begin{eqnarray}\label{MCeg1}
   \mathcal{M}^{(1)}_C=\left ( \begin{array}{cc}
             -Y^{(1/2)\over{q+1}} & ~-{1\over2}Y^{3\over{q+1}} \\
             {1\over2}\left(Y^{-{2\over{q+1}}}+1\right) & ~Y^{(3/2)\over{q+1}}
           \end{array} \right).
\end{eqnarray}
The power constants $\{a,b,c,d\}$ could as well be prescribed to be $q$-dependent. For example,
\begin{eqnarray}\label{MCeg2}
   \mathcal{M}^{(2)}_C=\left ( \begin{array}{cc}
             Y^{-{2\sinh^2(q)\over{q+1}}} & ~-{1\over2}Y^{4\cosh^2(q)\over{q+1}} \\
             {1\over2}\left(Y^{-{4\cosh(2q)\over{q+1}}}+1\right) & ~-Y^{2\cosh^2(q)\over{q+1}}
           \end{array} \right).
\end{eqnarray}
Many other sets of conditions on the constants exist such that condition (\ref{condhs}) is satisfied.

\subsubsection{Significance of Sub-Class C}\label{SignifC}

For any given order $n\ge4$, there exist sub-classes of pruned models where all of the matrix elements are composed of strictly distinct random variables chosen from the set $\left\{Y^{{p_1}\over{q+1}},Y^{{p_2}\over{q+1}},\ldots,Y^{{p_n}\over{q+1}}:n\in\mathbb{N}\right\}$ (where the real parameters $\left\{{p_1},{p_2},\ldots,{p_n}:n\in\mathbb{N}\right\}$ are all non-zero and distinct). The significance of sub-class C is that it is the simplest such sub-class (under the auxiliary condition that $\left\{a,b,c,d\right\}$ are all non-zero). 

Sub-class C is a special case of the more general sub-class of random matrices:
\begin{eqnarray}\label{MCmore}
   \mathcal{M}_G=\left ( \begin{array}{cc}
             g_1(X)+c_1+c_2Y^{a\over{q+1}} & c_3+c_4Y^{b\over{q+1}} \\
             c_5+c_6Y^{c\over{q+1}} & ~g_2(X)+c_7+c_8Y^{d\over{q+1}}
           \end{array} \right),
\end{eqnarray}
where there are no restrictions on the constants $\{c_j:j=1,\ldots,8\}$ beyond those required to satisfy condition (\ref{condhs}) and the power constants $\{a\neq1,b,c,d\neq1\}$ are restricted from being equal.\footnote{With reference to model (\ref{MBG1}), both $\mathcal{M}_C$ and $\mathcal{M}_G$ are sub-classes of class $\mathcal{M}_4$. The restrictions on the power constants given here as part of the definition of sub-class G guarantee that sub-classes A and G are disjoint.} As an example, one member of sub-class G is the random matrix:
\begin{eqnarray}\label{MCmoreEG}
   \mathcal{M}^{(1)}_G=\left ( \begin{array}{cc}
             \sec^{-1}(q|X|+1)-{\pi}Y^{-{2\over{q+1}}} & {1\over4}+Y^{-{4\over{q+1}}} \\
             {\pi^2\over4}\left(Y^{2\over{q+1}}-1\right) & ~-\csc^{-1}(q|X|+1)
           \end{array} \right),
\end{eqnarray}
which satisfies condition (\ref{condhs}) with $k={\pi^2\over4}$. Note that, in this case, each off-diagonal and lower-diagonal pair of elements is mutually independent. 

\section{Simple Symmetric Models}\label{RSegsSCs}

Model (\ref{MBG1}) can also generate symmetric sub-classes of models. The extra condition of symmetry (in addition to condition (\ref{condhs})) is actually non-trivial since its imposition introduces algebraic constraints that often result in sets of conditions that have no fully real solutions. It can also happen that satisfaction of condition (\ref{condhs}) precludes symmetry outright (e.g., sub-class B). Imposing symmetry reduces the number of distinct summand coefficients (i.e., the $\{k_m\}$ in model (\ref{MBG1}) or the constants $\{c_j\}$ in the context of sub-classes) and the reduction in algebraic degrees-of-freedom makes (non-trivial) satisfaction of condition (\ref{condhs}) more difficult. Satisfaction of condition (\ref{condhs}) can however be conveniently enforced by making use of a variable prefactor $\mathcal{A}$ that multiplies each of the matrix elements. (Recall that $\mathcal{A}=1$ for all sub-classes of models considered so far.) Any models that result from using such prefactors will (by definition) not be cropped, pruned, nor trimmed. 

\subsection{Simple Low-Order Models: Examples}

Among the simplest (non-trivial) models are the following symmetric sub-classes of $\mathcal{M}_1$ and $\mathcal{M}_2$ (respectively):
\begin{subequations}\label{RS1}  
\begin{eqnarray}\label{RS1a}
   {1\over\sqrt{c^2(q)Y^{-{2\over{q+1}}}+1}}\left ( \begin{array}{cc}
             c(q) & ~{1\over2}\left(c(q)+Y^{1\over{q+1}}\right) \\ [0.15cm]
             {1\over2}\left(c(q)+Y^{1\over{q+1}}\right) & Y^{1\over{q+1}}
           \end{array} \right),
\end{eqnarray}
\begin{eqnarray}\label{RS1b}
   {1\over\sqrt{c^2(q)Y^{-{2\over{q+1}}}+1}}\left ( \begin{array}{cc}
             c(q) & ~\sqrt{c(q)\over{2}}Y^{1\over{2(q+1)}} \\ [0.15cm]
             \sqrt{c(q)\over{2}}Y^{1\over{2(q+1)}} & Y^{1\over{q+1}}
           \end{array} \right),
\end{eqnarray}
\end{subequations}
where $c(q)$ is a real (possibly $q$-dependent) constant that can be arbitrarily prescribed. 

Models (\ref{RS1}) can be immediately generalized as follows:
\begin{subequations}\label{RS2} 
\begin{eqnarray}\label{RS2a}
   {1\over\sqrt{Y^{{2(a-1)\over{q+1}}}+Y^{{2(b-1)\over{q+1}}}}}\left ( \begin{array}{cc}
             Y^{a\over{q+1}} & ~{1\over2}\left(Y^{a\over{q+1}}+Y^{b\over{q+1}}\right) \\ [0.15cm]
             {1\over2}\left(Y^{a\over{q+1}}+Y^{b\over{q+1}}\right) & Y^{b\over{q+1}}
           \end{array} \right),~a \neq b 
\end{eqnarray}
\begin{eqnarray}\label{RS2b}
   {1\over\sqrt{Y^{{2(a-1)\over{q+1}}}+Y^{{2(b-1)\over{q+1}}}}}\left ( \begin{array}{cc}
             Y^{a\over{q+1}} & ~{1\over\sqrt{2}}Y^{a+b\over{2(q+1)}} \\ [0.15cm]
             {1\over\sqrt{2}}Y^{a+b\over{2(q+1)}}  & Y^{b\over{q+1}}
           \end{array} \right),~a \neq b 
\end{eqnarray}
\end{subequations}
where the strictly unequal (and independent) power constants $\{a,b\}$ are real (possibly $q$-dependent) constants that can be arbitrarily prescribed (e.g., $\{a=\pi,b=\ln(q+{1\over2})\}$). Models (\ref{RS2a}) and (\ref{RS2b}) are generally sub-classes of $\mathcal{M}_2$ and $\mathcal{M}_3$, respectively (the exceptions occur when one of $\{a,b\}$ is zero in which case (\ref{RS2a}) and (\ref{RS2b}) are sub-classes of $\mathcal{M}_1$ and $\mathcal{M}_2$, respectively). 

A related model with zero trace is:
\begin{eqnarray}\label{RS3}
   {1\over\sqrt{Y^{{a-2\over{q+1}}}+Y^{{b-2\over{q+1}}}}}\left ( \begin{array}{cc}
             Y^{a\over{2(q+1)}} & Y^{b\over{2(q+1)}} \\
             Y^{b\over{2(q+1)}} & ~-Y^{a\over{2(q+1)}}
           \end{array} \right),~a \neq b
\end{eqnarray}
which is a symmetric sub-class of $\mathcal{M}_2$ (unless one of $\{a,b\}$ is zero in which case (\ref{RS3}) is a sub-class of $\mathcal{M}_1$).

\subsection{Model Construction Using Preset Polynomial Functions of $Y^{1\over{q+1}}$}\label{polyconstructs}

There is no unique approach to constructing symmetric sub-classes of model (\ref{MBG1}). One could employ general algebraic approaches when considering the matrix-element summations (as was done in Section \ref{NonSymMods}), but one could also make use of specific predefined finite-sums with prescribed constants. To illustrate, consider (for simplicity) the case where the summand coefficients $\{k_m\}$ in model (\ref{MBG1}) are prescribed real numbers and the power constants $\left\{{p_1},{p_2},\ldots,{p_n}\right\}$ are positive integers; this corresponds to considering different polynomial functions of the fundamental random variable $Y^{1\over{q+1}}$. One could determine the general integer sets of values of $\left\{{p_1},{p_2},\ldots,{p_n}\right\}$ that satisfy condition (\ref{condhs}), which aligns with the general algebraic approach used in Section \ref{NonSymMods}, or one could alternatively prescribe the values of these constants (i.e., use specific preset polynomials) and then enforce the satisfaction of condition (\ref{condhs}) through use of the matrix prefactor $\mathcal{A}$. The latter approach will be employed in the following two subsections. 

The choice of polynomial(s) in the above-described construction is arbitrary. Polynomials obtained from binomial expansions are simple and well-suited due to their inherent symmetry. Sets of orthogonal polynomials are also convenient due to their many special properties and recurrence relations. Some specific examples are given in the following two subsections. 

\subsection{Polynomial Constructions 1: Binomial Expansions}

\subsubsection{Model 1} 

An example utilizing binomial expansions in the diagonal entries is the following general symmetric sub-class: 
\begin{eqnarray}\label{RSniceEG1}
\mathcal{A}\left ( \begin{array}{cc}
            \left(c(q)+Y^{1\over{q+1}}\right)^n & ~~\mathcal{P}\left(Y^{1\over{q+1}};c(q)\right) \\ [0.15cm]
             \mathcal{P}\left(Y^{1\over{q+1}};c(q)\right) & ~\left(c(q)-Y^{1\over{q+1}}\right)^n
           \end{array} \right), 
\end{eqnarray}
where $n$ is a positive integer, $\mathcal{A}$ is a variable prefactor, and $\mathcal{P}$ is a polynomial in the variable $Y^{1\over{q+1}}$ (with $c(q)$-dependent coefficients). $\left\{\mathcal{A},\mathcal{P}\right\}$ are non-unique (for each fixed value of $n$) and can be freely prescribed subject to condition (\ref{condhs}). 

\noindent \textbf{Example 1}: one sub-class when $n=3$ is:
\begin{subequations}\label{RSniceEG}
\begin{eqnarray}\label{RSniceEG2}
{1\over{\sqrt{9c^4+Y^{4\over{q+1}}}}}\left ( \begin{array}{cc}
            \left(c+Y^{1\over{q+1}}\right)^3 & ~~3c^2Y^{1\over{q+1}}-Y^{3\over{q+1}} \\ [0.15cm]
             3c^2Y^{1\over{q+1}}-Y^{3\over{q+1}} & ~\left(c-Y^{1\over{q+1}}\right)^3
           \end{array} \right). 
\end{eqnarray}

\noindent \textbf{Example 2}: one sub-class when $n=4$ is:
\begin{eqnarray}\label{RSniceEG3}
{1\over{c\sqrt{c^4+Y^{4\over{q+1}}}}}\left ( \begin{array}{cc}
            \left(c+Y^{1\over{q+1}}\right)^4 & ~~4c^3Y^{1\over{q+1}}-4cY^{3\over{q+1}} \\ [0.15cm]
             4c^3Y^{1\over{q+1}}-4cY^{3\over{q+1}} & ~\left(c-Y^{1\over{q+1}}\right)^4
           \end{array} \right).
\end{eqnarray}
\end{subequations}
Note that the degree of $\mathcal{P}$ in (\ref{RSniceEG1}) need not be equal to $n$. For instance, in Example 2 above (where $n=4$), $\mathcal{P}$ is a polynomial of degree $3$ (in the variable $Y^{1\over{q+1}}$). 

\subsubsection{Model 2} 

An example utilizing binomial expansions in only one of the diagonal entries are the symmetric sub-classes: 
\begin{eqnarray}\label{RSswapBiN}
\mathcal{A}_\pm\left ( \begin{array}{cc}
            \left(c(q) \pm Y^{1\over{q+1}}\right)^n & ~~\mathcal{P}_\pm\left(Y^{1\over{q+1}};c(q)\right) \\ [0.15cm]
             \mathcal{P}_\pm\left(Y^{1\over{q+1}};c(q)\right) & ~c(q)^n \pm Y^{n\over{q+1}}
           \end{array} \right), 
\end{eqnarray}
where (as in Model 1) $n$ is a positive integer, $\mathcal{A}_\pm$ are variable prefactors, and $\mathcal{P}_\pm$ are polynomials in the variable $Y^{1\over{q+1}}$. In each case (i.e., in the ``$+$'' or ``$-$'' case), $\left\{\mathcal{A}_\pm,\mathcal{P}_\pm\right\}$ are non-unique (for each fixed value of $n$) and can be freely prescribed subject to condition (\ref{condhs}). 

\noindent \textbf{Example 1}: one ``$+$'' sub-class when $n=3$ is:
\begin{subequations}\label{RSswapEGs}
\begin{eqnarray}\label{RSswapEG1}
{1\over{c\sqrt{c^2+Y^{2\over{q+1}}}}}\left ( \begin{array}{cc}
            \left(c+Y^{1\over{q+1}}\right)^3 & ~~{3\over2}c^2Y^{1\over{q+1}}-{3\over2}cY^{2\over{q+1}} \\ [0.15cm]
             {3\over2}c^2Y^{1\over{q+1}}-{3\over2}cY^{2\over{q+1}} & ~c^3+Y^{3\over{q+1}}
           \end{array} \right). 
\end{eqnarray}

\noindent \textbf{Example 2}: another ``$+$'' sub-class when $n=3$ is:
\begin{eqnarray}\label{RSswapEG2}
{1\over{c\left(c+Y^{1\over{q+1}}\right)}}\left ( \begin{array}{cc}
            \left(c+Y^{1\over{q+1}}\right)^3 & ~~{3\over2}c^2Y^{1\over{q+1}}+{3\over2}cY^{2\over{q+1}} \\ [0.15cm]
             {3\over2}c^2Y^{1\over{q+1}}+{3\over2}cY^{2\over{q+1}} & ~c^3+Y^{3\over{q+1}}
           \end{array} \right).
\end{eqnarray}
\end{subequations}

\subsubsection{Model 3}

An example having zero trace is the general symmetric sub-class (of $\mathcal{M}_n$): 
\begin{eqnarray}\label{RSgenEG}
\mathcal{A}_n\left ( \begin{array}{cc}
            \left(c(q)-Y^{1\over{q+1}}\right)^n & ~~\left(c(q)+Y^{1\over{q+1}}\right)^n \\ [0.15cm]
             \left(c(q)+Y^{1\over{q+1}}\right)^n & ~-\left(c(q)-Y^{1\over{q+1}}\right)^n
           \end{array} \right), 
\end{eqnarray}
where $n$ is a positive integer. 

\noindent \textbf{Example}: For $n=\{1,2,3\}$, the matrix prefactor $\mathcal{A}_n$ is:
\begin{subequations}\label{RSgenEGtheDs}
\begin{eqnarray}
&&\mathcal{A}_1={1\over{2\sqrt{c^2Y^{-{2\over{q+1}}}+1}}}, \\
&&\mathcal{A}_2={1\over{2\sqrt{c^4Y^{-{2\over{q+1}}}+6c^2+Y^{2\over{q+1}}}}}, \\
&&\mathcal{A}_3={1\over{2\sqrt{c^6Y^{-{2\over{q+1}}}+15c^4+15c^2Y^{2\over{q+1}}+Y^{4\over{q+1}}}}}.
\end{eqnarray}
\end{subequations}

\subsubsection{Mixed Constructions}

In constructing models using binomial expansions, it is not necessary for all matrix elements to be strictly comprised of only polynomial functions of $Y^{1\over{q+1}}$. The following are two simple low-order examples. 

\noindent \textbf{Example 1}: 
\begin{subequations}\label{RSLO}
\begin{eqnarray}\label{RSLO1}
   {1\over\sqrt{c^2+Y^{{2\over{q+1}}}}}\left ( \begin{array}{cc}
             \left(c-Y^{1\over{q+1}}\right)^2 & ~\sqrt{2c}Y^{3\over{2(q+1)}} \\ [0.15cm]
             \sqrt{2c}Y^{3\over{2(q+1)}} & c^2-Y^{2\over{q+1}}
           \end{array} \right),
\end{eqnarray}
which is a sub-class of $\mathcal{M}_3$. 

\noindent \textbf{Example 2}: 
\begin{eqnarray}\label{RSLO2}
   {1\over{c}\sqrt{c^2+Y^{{2\over{q+1}}}}}\left ( \begin{array}{cc}
             \left(c-Y^{1\over{q+1}}\right)^3 & ~{3\over\sqrt{2}}\left(cY^{1\over{q+1}}\right)^{3/2} \\ [0.15cm]
            {3\over\sqrt{2}}\left(cY^{1\over{q+1}}\right)^{3/2} & c^3-Y^{3\over{q+1}}
           \end{array} \right),
\end{eqnarray}
which is a sub-class of $\mathcal{M}_4$.
\end{subequations}

\noindent Note that (ignoring matrix prefactors) the off-diagonal elements of models (\ref{RSLO1}) and (\ref{RSLO2}) are (apart from a constant) comprised of the same non-polynomial function of $Y^{1\over{q+1}}$ while their diagonal elements are comprised of (different) polynomial functions of $Y^{1\over{q+1}}$. 

\subsubsection{Non-Polynomial Generalization}

The binomial expansions $\left(c(q) \pm Y^{1\over{q+1}}\right)^n$ were used in models (\ref{RSniceEG1}), (\ref{RSswapBiN}), (\ref{RSgenEG}), and (\ref{RSLO}). One could also use the more general binomial expansions $\left(Y^{a\over{q+1}} \pm Y^{b\over{q+1}}\right)^n$, where $a \neq b$ are arbitrary real parameters. This would yield non-polynomial generalizations of the preceding constructions. 

\noindent \textbf{Example}: Defining $C(a,b)\equiv\sqrt{Y^{2a\over{q+1}}+Y^{{2b\over{q+1}}}}$, the resulting generalizations of models (\ref{RSLO1}) and (\ref{RSLO2}) would be (respectively):
\begin{subequations}\label{RSLOGEN}
\begin{eqnarray}\label{RSLOGEN1}
   {1\over{C(a,b)Y^{b-1\over{q+1}}}}\left ( \begin{array}{cc}
             \left(Y^{a\over{q+1}}-Y^{b\over{q+1}}\right)^2 & ~\sqrt{2}Y^{a+3b\over{2(q+1)}} \\ [0.15cm]
             \sqrt{2}Y^{a+3b\over{2(q+1)}} & ~Y^{2a\over{q+1}}-Y^{2b\over{q+1}}
           \end{array} \right),
\end{eqnarray}
which is a sub-class of $\mathcal{M}_4$, and 
\begin{eqnarray}\label{RSLOGEN2}
   {1\over{C(a,b)Y^{a+b-1\over{q+1}}}}\left ( \begin{array}{cc}
             \left(Y^{a\over{q+1}}-Y^{b\over{q+1}}\right)^3 & ~{3\over\sqrt{2}}Y^{3(a+b)\over{2(q+1)}} \\ [0.15cm]
            {3\over\sqrt{2}}Y^{3(a+b)\over{2(q+1)}} & ~Y^{3a\over{q+1}}-Y^{3b\over{q+1}}
           \end{array} \right),
\end{eqnarray}
which is a sub-class of $\mathcal{M}_5$.
\end{subequations}

\subsection{Polynomial Constructions 2: Orthogonal Polynomials}

As mentioned in Section \ref{polyconstructs}, model construction using predefined polynomials could also employ sets of orthogonal polynomials, including (but not limited to) Legendre, Hermite, Laguerre, and Chebyshev polynomials. Examples making use of the latter two polynomial sets are given below. In the following, the random variable $W\equiv Y^{1\over{q+1}}$ [c.f., Eq.~(\ref{RayisGG})]. 

\subsubsection{Model 1: Laguerre Polynomials}

Let $\mathcal{L}_n$ denote the Laguerre polynomial of degree $n$, where $n=\{0,1,2,3,\ldots\}$. The following model (among others) is obtained using the fundamental recurrence relation for Laguerre polynomials:
\begin{eqnarray}\label{RSLAGGYEG}
{1\over{\sqrt{1+(2n+1)^2W^{-2}}\mathcal{L}_n(W)}}\left ( \begin{array}{cc}
            -\mathcal{L}_{n+1}(W) & ~~{1\over2}(W+2n+1)\mathcal{L}_n(W) \\ [0.05cm]
             {1\over2}(W+2n+1)\mathcal{L}_n(W) & ~n^2\mathcal{L}_{n-1}(W)
           \end{array} \right).
\end{eqnarray}

\noindent \textbf{Example}: The case $n=2$ yields the model
\begin{subequations}\label{RSLAGGYEG1}
\begin{eqnarray}\label{RSLAGGYEG1a}
\mathcal{A}_{\mathcal{L}_2}(W) \left ( \begin{array}{cc}
            W^3-9W^2+18W-6 & ~~{1\over2}W^3+{1\over2}W^2-9W+5 \\ [0.05cm]
             {1\over2}W^3+{1\over2}W^2-9W+5 & ~-4W+4
           \end{array} \right),
\end{eqnarray}
where
\begin{equation}\label{RSLAGGYEG1b}
\mathcal{A}_{\mathcal{L}_2}(W)={1\over{\sqrt{1+25W^{-2}}\left(W^2-4W+2\right)}}.
\end{equation}
\end{subequations}

\subsubsection{Model 2: Chebyshev Polynomials}

Let $\{\mathcal{T}_n,\mathcal{U}_n\}$ denote the Chebyshev polynomials of the first and second kinds (respectively) of degree $n$. The recurrence relations for each of these two polynomial sets and the various interrelationships between them allow the construction of many different symmetric models. For instance, one model is: 
\begin{eqnarray}\label{RSTEBBYEG}
{1\over{\left(W+W^{-1}\right)\mathcal{U}_{n-1}(W)}}\left ( \begin{array}{cc}
            W\mathcal{T}_{n}(W) & ~~W\mathcal{U}_{n-1}(W) \\ [0.03cm]
             W\mathcal{U}_{n-1}(W) & ~\mathcal{T}_{n+1}(W)
           \end{array} \right).
\end{eqnarray}

\noindent \textbf{Example}: The case $n=2$ yields the model
\begin{eqnarray}\label{RSTEBBYEG1}
{1\over{2\left(W^2+1\right)}}\left ( \begin{array}{cc}
            2W^3-W & ~2W^2 \\ [0.03cm]
             2W^2 & ~~4W^3-3W
           \end{array} \right).
\end{eqnarray}
 
\section{Complex Matrices and Eigenvalues}\label{cmplxySEC}

Model (\ref{MBG1}) admits complex matrices having real or complex eigenvalues. As long as condition (\ref{condhs}) is satisfied, the Brody distribution holds for complex matrices and eigenvalues as well. An exhaustive treatment of all six possible complex cases will not be given here. For simplicity, only the two cases where the trace of $\mathcal{M}_n$ is real will be discussed. For purposes of illustration, it suffices to consider the complex generalization of (the non-symmetric) sub-classes A and B and to give simple examples chosen from those sub-classes.

\subsection{Complex Matrices with Real Eigenvalues}\label{cmplxy}

Condition (\ref{MB2}) for sub-class A, conditions (\ref{MBtunies1})-(\ref{MBtunies3}) for sub-class B, and conditions (\ref{MCegGen1a})-(\ref{MCegGen1b}) for sub-class C, each constitute an underdetermined quadratic polynomial system. As previously discussed, these systems possess an infinite number of real and complex solutions. In defining sub-classes A, B, and C, the constants $\{c_j\}$ were stipulated to be real. This was done to simplify the conditioning of the constants, but is not a strict requirement. If the constants $\{c_j\}$ are permitted to be complex, then the defining conditions can be generalized so as to admit complex matrices with real eigenvalues. For example, condition (\ref{MB2}) for sub-class A would be replaced by the following three conditions: 
\begin{subequations}\label{condsModAcmplx}
\begin{eqnarray}
&&\text{Im}\left\{c_1+c_4\right\}=0, \label{Acmplx1} \\
&&\text{Re}\left\{(c_1-c_4)^2+4c_2c_3\right\}>0, \label {Acmplx2} \\
&&\text{Im}\left\{(c_1-c_4)^2+4c_2c_3\right\}=0. \label{Acmplx3}
\end{eqnarray}
\end{subequations}
An example satisfying conditions (\ref{condsModAcmplx}) is:
\begin{eqnarray}\label{MBegcmplx}
   \mathcal{M}^{\text{nH}(1)}_A=\left ( \begin{array}{cc}
             X+(3-i)Y^{1\over{q+1}} & \left(2+{1\over2}i\right)Y^{-1} \\
             \left(2+{1\over2}i\right)Y^{q+3\over{q+1}} & X+(1+i)Y^{1\over{q+1}}
           \end{array} \right),
\end{eqnarray}
which is non-Hermitian. Note that the power constants $\{a,b\}$ were again chosen to be $q$-dependent and were prescribed as follows: $\{a(q)=-(q+1),b(q)=q+3\}$. Complex Hermitian matrices are precluded from sub-class A due to the restriction $a \neq b$. 

The conditions for sub-class B can be similarly generalized but the details will be omitted here. It can be shown that the conditions in (\ref{MBtunies1})-(\ref{MBtunies3}) preclude the existence of Hermitian matrices for sub-class B. It is interesting to give one example of a complex sub-class B random matrix that belongs to a class of random matrices that has (to the author's knowledge) received no attention in RMT: complex symmetric random matrices having real eigenvalues.\footnote{A complex symmetric matrix is defined to be a square matrix with complex entries that is equal to its transpose (and is thus non-Hermitian). General complex symmetric random matrices having complex eigenvalues have (unlike those with exclusively real eigenvalues) received considerable attention in recent years (see, for example, Refs.~\cite{Indys,Germys} and references therein).} In general, complex symmetric matrices have complex diagonal and off-diagonal entries, in which case, the eigenvalues will be generally complex. If however the trace is real, then the eigenvalues can be real or complex conjugates (depending on the sign of the discriminant). A curious feature of sub-class B is that the conditions associated with its definition do not allow for real symmetric matrices but they do allow for complex symmetric matrices. A simple example that derives from Case II and whose eigenvalues are real is the following:
\begin{eqnarray}\label{ModBcomplxsym}
   \mathcal{M}^{\text{cs(1)}}_B=\left ( \begin{array}{cc}
             Y^{a\over{q+1}} & {i\over2}\left(Y^{a\over{q+1}}-Y^{b\over{q+1}}\right) \\
             {i\over2}\left(Y^{a\over{q+1}}-Y^{b\over{q+1}}\right) & -Y^{b\over{q+1}}
           \end{array} \right),
\end{eqnarray}
where the (real) power constants $\{a,b\}$ are unequal and satisfy condition (\ref{MBtunies2}). 

\subsection{Complex-Conjugate Eigenvalues}\label{CCevsSec}

In the present context, complex-conjugate (CC) eigenvalues occur when two conditions are simultaneously met: (i) $\text{Tr}(\mathcal{M}_n)$ is real; and (ii) the discriminant constant $k$ in Eq.~(\ref{condhs}) is real and negative. As a consequence of condition (ii), $\sqrt{\mathcal{D}(\mathcal{M}_n)}$ is pure imaginary. The definition of the spacing given by Eq.~(\ref{spacsG}) applies only when the eigenvalues are real and requires a minor modification. For any two complex eigenvalues, the spacing between them is the Euclidean distance in the complex plane. However, since CC eigenvalues have equal real parts, the standard two-dimensional Euclidean distance formula reduces such that the spacing is simply the linear distance between their (opposite and equal) imaginary parts:
\begin{equation}\label{spacsCC}
S\equiv\left|\lambda_{+}-\lambda_{-}\right|=2\left|\text{Im}(\lambda_{\pm})\right|.
\end{equation} 
Then, under condition (\ref{condhs}) with $k<0$, the random spacing between CC eigenvalues is:
\begin{equation}\label{spacsCCF}
S=\left|\sqrt{\mathcal{D}(\mathcal{M}_n)}\right|=\text{Im}\left(\sqrt{k}\right)Y^{1\over{q+1}}=\sqrt{|k|}Y^{1\over{q+1}}.
\end{equation}
Comparing (\ref{spacsCCF}) and the equivalent result (\ref{spacs}) for real eigenvalues, it is clear that the spacing distribution for CC eigenvalues will again be the Brody distribution. The derivation in Section \ref{NNSD} proceeds exactly as before with the $\sqrt{k}$ term replaced by $\sqrt{|k|}$; the final result is unchanged. 

Some illustrative examples having CC eigenvalues are given below.

\noindent \textbf{Example 1}: Matrix $\mathcal{M}^{(1)}_{A}$ [Eq.~(\ref{MBnsEG})] with constants $\{c_1=2,c_2=1,c_3=-1,c_4=1\}$
is a real asymmetric matrix with $k=-{3}$. 

\noindent \textbf{Example 2}: $\mathcal{M}^{(1)}_{A}$ [Eq.~(\ref{MBnsEG})] with constants $\{c_1=1+2i,c_2=1+i,c_3=1-i,c_4=1-2i\}$
is a complex asymmetric non-Hermitian matrix with $k=-{8}$.

\noindent \textbf{Example 3}: 
\begin{eqnarray}\label{ModBcomplxsym2}
   \mathcal{M}^{\text{cs(2)}}_B=\left ( \begin{array}{cc}
             Y^{a\over{q+1}} & {i\over2}\left(Y^{a\over{q+1}}+Y^{b\over{q+1}}\right) \\
             {i\over2}\left(Y^{a\over{q+1}}+Y^{b\over{q+1}}\right) & Y^{b\over{q+1}}
           \end{array} \right),
\end{eqnarray}
where the power constants $\{a,b\}$ are unequal and satisfy condition (\ref{MBtunies2}), is a complex symmetric matrix with $k=-{4}$. 

\section{Generalizations and Extensions}\label{GensExts}

\subsection{Generalization of the Exponents of $Y$}\label{GenExps}

In model (\ref{MBG1}), the set $\left\{{p_1},{p_2},\ldots,{p_n}:n\in\mathbb{N}\right\}$ were defined to be a constrained set of real constants that could have parametric dependencies. In the examples provided for sub-classes A and B, the constrained constants $\{a,b\}\equiv\{p_1,p_2\}$ were prescribed to be $q$-dependent constants (i.e., constants whose values depend on the value of $q$). While this prescription provides a useful generalization of the models in terms of the parameter $q$, it is not required. Condition (\ref{MB3}) for sub-class A and condition (\ref{MBtunies2}) for Case II of sub-class B are sufficiently general so as to allow $\{a,b\}$ to be any set of constants, parameters, or even variables that satisfy the condition in question. An example for each of the aforementioned possibilities is given below.

\subsubsection{Illustrative Examples}\label{GenExpsEgs}

\begin{subequations}\label{EXPegs} 
\noindent \textbf{Example 1} (Constants):
\begin{eqnarray}\label{ConstsEG}
a=J_1\left({1\over\pi}\right)N_0\left({1\over\pi}\right),~b=-J_0\left({1\over\pi}\right)N_1\left({1\over\pi}\right), 
\end{eqnarray}
where $\{J_p,N_p\}$ are Bessel functions of the first and second kinds (respectively) of order $p$.  

\noindent \textbf{Example 2} (Parameters):
\begin{eqnarray}\label{ParamsEG}
a(\ell,p)=\sin^2\left({\ell\pi\over2}\right)+\sin^2(p),~b(\ell,p)=\cos^2\left({\ell\pi\over2}\right)+\cos^2(p),~~~\ell\in\mathbb{Z},~p\in\mathbb{R}. 
\end{eqnarray}

\noindent \textbf{Example 3} (Random Variables):
\begin{eqnarray}\label{RVsEG}
a(U,V)=\text{sech}^2(U)+\coth^2(V),~b(U,V)=\tanh^2(U)-\text{csch}^2(V), 
\end{eqnarray}
where the random variables $\{U,V\}$ are mutually independent (and also independent of $Y$) and can have any continuous distributions $f_U(u)$ and $f_V(v)$, respectively.\footnote{In theory, the support of $V$ should not include zero.}
\end{subequations} 
For sub-class A, using (\ref{RVsEG}) yields:
\begin{eqnarray}\label{MBnsEGRVs}
   \mathcal{M}^{(3)}_{A}=\left ( \begin{array}{cc}
             X+c_1Y^{1\over{q+1}} & ~c_2Y^{{{\text{sech}^2(U)+\coth^2(V)}\over{q+1}}} \\
             c_3Y^{{{\tanh^2(U)-\text{csch}^2(V)}\over{q+1}}} & X+c_4Y^{1\over{q+1}}
           \end{array} \right),
\end{eqnarray}
where it has been assumed that constants $\{c_2,c_3\}$ are both non-zero.

While the above discussion focussed on sub-classes A and B, the same comments of course apply to all pertinent sub-classes of models (real or complex) considered in Sections \ref{NonSymMods}-\ref{cmplxySEC}. For example, the random matrix 
\begin{eqnarray}\label{MCegRV}
   \mathcal{M}^{(3)}_C=\left ( \begin{array}{cc}
             Y^{{2\sin^2(V)\over{q+1}}} & ~-{1\over2}Y^{4\cos^2(V)\over{q+1}} \\
             {1\over2}\left(Y^{-{4\cos(2V)\over{q+1}}}+1\right) & ~-Y^{2\cos^2(V)\over{q+1}}
           \end{array} \right),
\end{eqnarray}
where the random variable $V$ is independent of $Y$ and can have any arbitrary continuous distribution $f_V(v)$, also satisfies conditions (\ref{MCegGen}) of sub-class C. With respect to the real-symmetric models (\ref{RS2},\ref{RS3},\ref{RSLOGEN}), the parameters $\{a,b\}$ are actually independent and so one could (if desired) set them as independent random variables (e.g., $\{a=\sin(U),b=\cos(V)\}$). 

In general, the numerators of the exponents of $Y$ in model (\ref{MBG1}) (i.e., the set $\left\{{p_1},{p_2},\ldots,{p_n}:n\in\mathbb{N}\right\}$) can be any set of constants, parameters, or random variables that satisfy condition (\ref{condhs}).

\subsubsection*{Comment: $q$-Independent Exponents}

For any sub-class of model (\ref{MBG1}) with $n\ge2$, it is possible to prescribe $q$-dependent values for parameters $\left\{{p_1},{p_2},\ldots,{p_n}:n\in\mathbb{N}\right\}$ in such manner that at least one exponent (of $Y$) is independent of $q$. For example, in the case of sub-class C with conditions (\ref{MCegGen}), prescribing the power constant $d=p(q+1)$, where $p\in\mathbb{R}\backslash\left[{1\over3},{2\over3}\right]$ is a real $q$-independent parameter, yields two $q$-independent exponents. As a specific example, choosing $p=2$ and constants $\{c_j\}$ as in model (\ref{MCeg2}) yields:
\begin{eqnarray}\label{MCegNOqs}
   \mathcal{M}^{(4)}_C=\left ( \begin{array}{cc}
             Y^{-{2q\over{q+1}}} & ~-{1\over2}Y^{4} \\
             {1\over2}\left(Y^{-{4(2q+1)\over{q+1}}}+1\right) & ~-Y^{2}
           \end{array} \right).
\end{eqnarray}
For another specific example, see model (\ref{MBegcmplx}). 

\subsection{Generalization of the Summand Coefficients $\{k_m\}$ in Model (\ref{MBG1})}\label{GenConsts}

For given $n\in\mathbb{N}$ in model (\ref{MBG1}), the set $\{k_m:m=1,2,3,\ldots,4n+4\}$ was defined to be a constrained set of (real or complex) constants. These coefficients are constrained in the sense that their values are restricted such that condition (\ref{condhs}) is satisfied, but they are otherwise arbitrary. In particular, these coefficients need not be fixed numbers; they could depend on the Brody parameter $q$ (or other user-defined parameters) or could even be random variables. 

\subsubsection{Illustrative Example}\label{GenConstsEg}

As an illustrative example, consider the coefficients $\{c_j:j=1,\ldots,6\}$ of sub-class C, which in a previous example (c.f., Section \ref{SubCegs}) were stipulated to be real constants satisfying conditions (\ref{MCegGen1a})-(\ref{MCegGen1b}). Setting 
\begin{eqnarray}
c_1=\widetilde{c_1}p(V),~c_4={\widetilde{c_4}\over p^2(V)},~c_5=\widetilde{c_5}p^2(V),~c_6={\widetilde{c_6}\over p(V)}
\end{eqnarray}
in (\ref{MCegGen1a})-(\ref{MCegGen1b}) yields the transformed conditions
\begin{subequations}\label{MCegGenConsts}
\begin{eqnarray}
\left(\widetilde{c_1}^2+4c_3\widetilde{c_5}\right)p^2(V)=0,~c_2=0, \\
{\left(\widetilde{c_6}^2+4c_3\widetilde{c_4}\right) \over p^2(V)}=0,~\widetilde{c_1}\widetilde{c_6}<0,
\end{eqnarray}
\end{subequations}
where the random variable $V$ is independent of $Y$ and can have any arbitrary continuous distribution $f_V(v)$, and $p(V)$ is any real function of $V$ having the property $p(V)\neq0$ for all supported values of $V$. For any given $p(V)$, system (\ref{MCegGenConsts}) again constitutes an underdetermined quadratic polynomial system possessing an infinite number of real (and complex) solutions. For example, one solution to system (\ref{MCegGenConsts}) is $\{\widetilde{c_1}=-1,c_2=0,c_3=-{1\over2},\widetilde{c_4}=\widetilde{c_5}={1\over2},\widetilde{c_6}=1\}$. Choosing power constants $\{a=0,b=4,c=-4,d=2\}$ (which satisfy conditions (\ref{MCegGen2})) and $p(V)=\cosh(qV)$ then yields:
\begin{eqnarray}\label{MCeg5}
   \mathcal{M}^{(5)}_C=\left ( \begin{array}{cc}
             -\cosh(qV) & ~-{1\over2}Y^{4\over{q+1}} \\
             {1\over2}\left(\cosh^2(qV)Y^{-{4\over{q+1}}}+\text{sech}^2(qV)\right) & ~\text{sech}(qV)Y^{2\over{q+1}}
           \end{array} \right).
\end{eqnarray}
Note that in the above example the upper diagonal and upper off-diagonal matrix elements are mutually independent.

\subsection{Cross-Over Transitions Between Other RMT Universality Classes}

The random variable $Y$ in model (\ref{MBG1}) was stipulated to have an exponential distribution [c.f., Eq.~(\ref{rayden})], which (subject to condition (\ref{condhs})) results in the eigenvalue spacing distribution exhibiting a cross-over transition between the Poisson and Wigner distributions as the parameter $q$ is varied between $0$ and $1$, respectively. The distribution of $Y$ can be changed, in which case, the spacing statistics for model (\ref{MBG1}) will undergo a cross-over transition between a different pair (or set) of distributions that may be of interest. 

An interesting example is the case where $Y\sim\text{Gamma}(\sigma_\text{g},\xi=2)$, that is, when $Y$ has a gamma distribution with shape parameter $\xi=2$ and arbitrary scale parameter $\sigma_\text{g}>0$:
\begin{equation}\label{gammaden}
f_Y(y;\sigma_\text{g},\xi=2)=\left({1\over\sigma_\text{g}^2}\right)y\exp\left(-{y\over\sigma_\text{g}}\right),~y>0.
\end{equation}
Note that $Y$ has a ``semi-Poisson'' distribution (in the density variable $y$) when $\sigma_\text{g}=1/2$. It can be shown (see Appendix \ref{sPtoGinPf}) that when the distribution of $Y$ is given by Eq.~(\ref{gammaden}) and condition (\ref{condhs}) holds, then the (mean-scaled) spacing distribution for model (\ref{MBG1}) is:
\begin{subequations}\label{Eqn:BrodySakhr} 
\begin{equation}\label{BrodySakhrp1}
P_{B}^{(II)}(z;q) = \alpha^2 (q+1) z^\beta\exp\left(-\alpha z^{q+1}\right), 
\end{equation}  
where 
\begin{equation}\label{BrodySakhrp2}
\alpha\equiv\alpha(q)=\left[ \Gamma \left( \frac{2q+3}{q+1}\right) \right]^{q+1}, 
\end{equation}
and
\begin{equation}\label{BrodySakhrp3}
\beta\equiv\beta(q)=2q+1.
\end{equation}
\end{subequations}
Distribution (\ref{Eqn:BrodySakhr}), which in the present paper will be referred to as ``Brody-II'', reduces to the semi-Poisson distribution
\begin{equation}\label{semiPoisson}
P_{sP}(z)=4z\exp(-2z)
\end{equation}
when $q=0$, and the Ginibre distribution
\begin{equation}\label{Ginibre}
P_{G}(z)={3^4\pi^2\over2^7}z^3
\exp\left(-{3^2\pi\over2^4}z^2\right) 
\end{equation} 
when $q=1$. Thus, in this case, model (\ref{MBG1}) exhibits a cross-over transition between semi-Poisson and Ginibre statistics (i.e., a transition between \emph{linear} and \emph{cubic} level-repulsion) as the parameter $q$ is varied between $0$ and $1$, respectively. For this transition, the level-repulsion exponent $\beta(q)$ (\ref{BrodySakhrp3}) quantifies the fractional degree of level repulsion between the linear and cubic endpoint cases. The semi-Poisson distribution is fundamentally relevant in many physical contexts (e.g., \cite{beyondsemiPs,Bogo2,JainpapsM2,Pbills2}) as is the Ginibre distribution (e.g., \cite{GrobeHaake,MPWQCD}); the latter is the NNSD pertinent to complex eigenvalues from Ginibre's ensemble of $2\times2$ complex non-Hermitian random matrices \cite{Ginib}. The occurrence of the Ginibre distribution in the context of model (\ref{MBG1}) is interesting since neither the matrices nor the eigenvalues need be complex. 
 
\section{Numerics}\label{NumEgs}

\subsection{Procedures}\label{procedsEGs}

To generate numerical realizations of any random-matrix model that derives from model (\ref{MBG1}), it is necessary to generate exponential variates (i.e., to generate sample numbers at random from an exponential distribution). A simple and widely-used method for generating variates from any probability distribution is the inverse transform sampling method (see, for example, Ref.~\cite{Norman}). In practical terms, the end result of applying the method in the exponential case amounts to two steps: first generate a random variate $U_o$ drawn from $\text{Uniform}(0,1)$ and then generate an exponential variate using the inverse transformation: $Y_o=-\sigma_\text{e} \ln(U_o)$. Using this method, it is thus only necessary to generate variates from the uniform distribution on the unit interval $(0,1)$, which can be easily accomplished using any of the standard numerical software packages. 

Armed with the capacity to generate sample matrices, it is then simply a matter of finding the corresponding eigenvalues and computing their spacings. The procedure for computing the sample set of spacings is as follows. Suppose that $N$ sample matrices are generated from the pertinent model. For the $n$th sample matrix ($n=1,\ldots,N$), the two real eigenvalues $\{\lambda_1,\lambda_2\}_n\equiv\Lambda_n$ are first numerically determined and their spacing
\begin{subequations}\label{spacsNUM}
\begin{equation}\label{realSs}
s_n\equiv\text{max}(\Lambda_n)-\text{min}(\Lambda_n)
\end{equation}
is then computed. When the two eigenvalues are not real but instead are complex conjugates, the above spacing formula is replaced by
\begin{equation}\label{spacsCCnums}
s_n\equiv\left|\lambda_{1}-\lambda_{2}\right|_n=2\left|\text{Im}(\lambda_{1/2})\right|_n.
\end{equation}
\end{subequations} 
The complete sample set of spacings is the set
\begin{subequations}\label{theSs} 
\begin{equation}
\mathbb{S}\equiv\left\{s_n:n=1,\ldots, N\right\}. 
\end{equation}
The sample set of mean-scaled spacings
\begin{equation}\label{scaledSsA}
{\mathbb{Z}}\equiv\left\{{z}_n \equiv s_n / \bar{s}:n=1,\ldots, N\right\}=\left({1\over\bar{s}}\right)\mathbb{S}, 
\end{equation}
where 
\begin{equation}\label{scaledSsB}
\bar{s}\equiv\bar{s}(N)={1\over{N}}\sum_{n=1}^{N} s_n
\end{equation}
\end{subequations}
is the sample mean spacing, is then computed. The sample spacing distribution, denoted here by $\hat{P}(z)$, is obtained by constructing a density histogram of the mean-scaled spacings $\mathbb{Z}$. This sample distribution (of set $\mathbb{Z}$) is then to be compared with the corresponding theoretical mean-scaled spacing distribution (i.e., the Brody distribution). Note that the ``unfolding'' procedure \cite{Haake} that is usually employed in statistical analyses of large random matrices has not been employed here. As discussed in Ref.~\cite{BerryAGN}, this procedure is not necessary in the case of $2\times2$ matrices. Notwithstanding this fact, dividing by the mean spacing is effectively a simple form of unfolding. 

In the following numerical examples, the number of realizations is set at $N=10^6$. The (positive) value of the scale parameter $\sigma_\text{e}$ is not however the same in all cases and will be specified with each example. For convenience, the random variable $X$ (whenever pertinent) is set as a standard normal. The MATLAB software package is employed for all numerics.

\subsection{Numerical Examples}

\subsubsection{Example 1: Model $\mathcal{M}^{(3)}_{A}$ [Eq.~(\ref{MBnsEGRVs})]}

\begin{figure}[h]
\vspace*{0.25cm} 
%\centering
\hspace*{-0.29cm}
\scalebox{0.62}{\includegraphics*{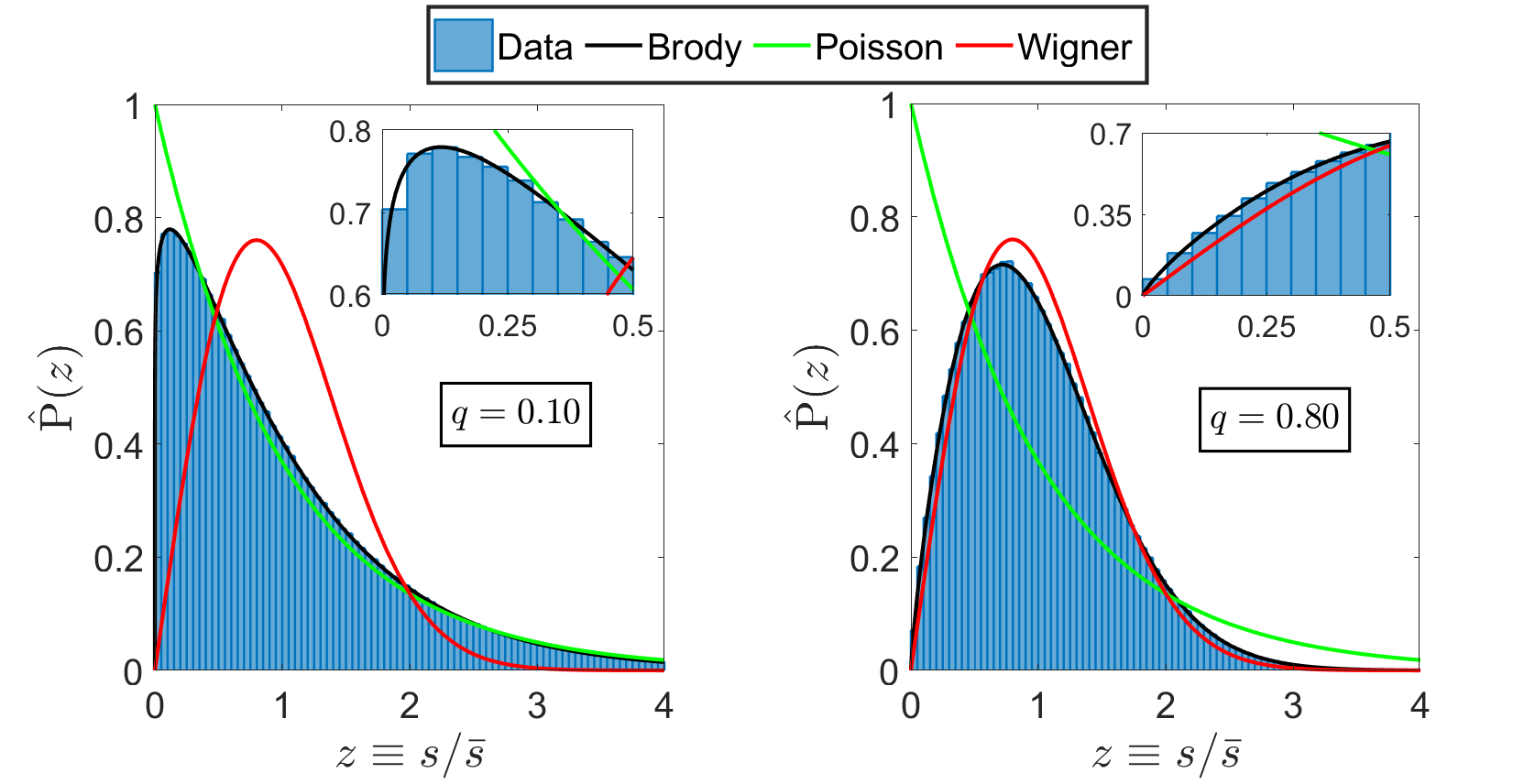}}
%\vspace*{-2.7cm}
\caption{\label{EG1} Eigenvalue spacing distributions for $N=10^6$ realizations of model $\mathcal{M}^{(3)}_{A}$ [Eq.~(\ref{MBnsEGRVs})]; values of the Brody parameter $q$ are as indicated. Insets show closer views at small spacings.}
\end{figure}

Applying the procedures of Section \ref{procedsEGs} to model $\mathcal{M}^{(3)}_{A}$ [Eq.~(\ref{MBnsEGRVs})], which is real and non-symmetric, yields the results shown in Fig.~\ref{EG1}; two different values of the Brody parameter $q$ (as indicated) are considered. The following parameter values and variables pertinent to model $\mathcal{M}^{(3)}_{A}$ were employed in the calculations: $\sigma_\text{e}=\ln(2)<1$, $\{c_1=c_2=c_3=1,c_4=-1\}$, $U\sim\text{Rayleigh}(\sigma_R=2)$, and $V\sim\text{THN}(\mu_{\text{h}}=1,\sigma_{\text{h}}^2=9)$, that is, $V$ has a truncated half-normal distribution with location parameter $\mu_{\text{h}}=1$, scale parameter $\sigma_{\text{h}}=3$, and (truncated) support $v\in[1,\infty)$. Numerics and theory are clearly consistent. 
 
\subsubsection{Example 2: Model $\mathcal{M}^{\text{nH}(1)}_A$ [Eq.~(\ref{MBegcmplx})]}

\begin{figure}[h]
\vspace*{0.25cm} 
%\centering
\hspace*{-0.29cm}
\scalebox{0.62}{\includegraphics*{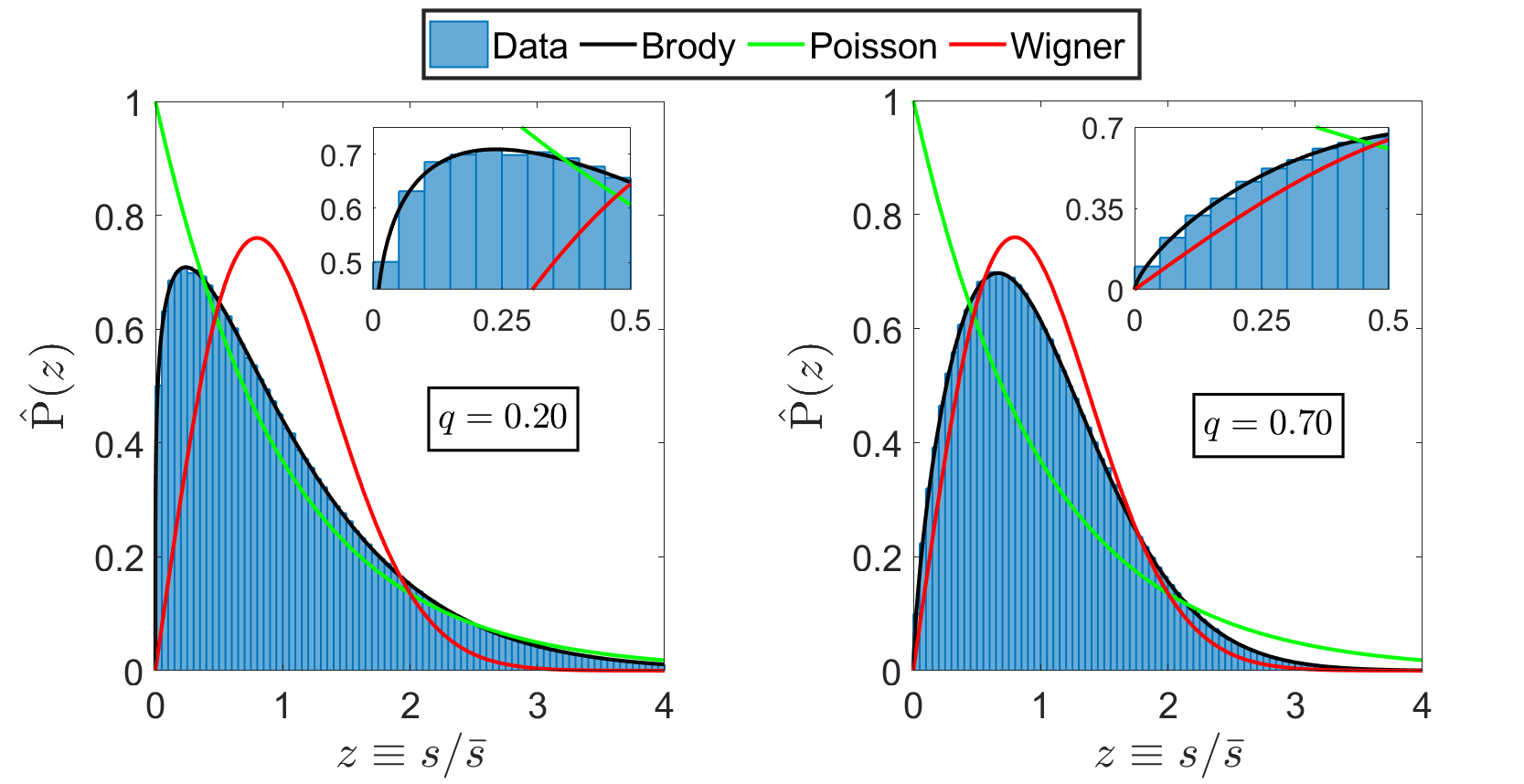}}
%\vspace*{-2.7cm}
\caption{\label{EG2} Eigenvalue spacing distributions for $N=10^6$ realizations of model $\mathcal{M}^{\text{nH}(1)}_A$ [Eq.~(\ref{MBegcmplx})]; values of the Brody parameter $q$ are as indicated. Insets show closer views at small spacings.}
\end{figure}

Results for model $\mathcal{M}^{\text{nH}(1)}_A$ [Eq.~(\ref{MBegcmplx})], which is complex and non-Hermitian, are shown in Fig.~\ref{EG2}; values of the Brody parameter $q$ are as indicated and the value of the scale parameter $\sigma_\text{e}=\sqrt{5}>1$. Numerics and theory are again consistent. 

It should be noted that for this model the numerical eigenvalues (as determined by MATLAB's numerical eigensolver) had tiny imaginary parts (magnitude $<10^{-13}$). These occur as a result of  numerical round-off errors. As a numerical fix, $\Lambda_n$ in Eq.~(\ref{realSs}) was replaced by Re$\{\Lambda_n\}$ thereby dropping these spurious imaginary parts. 

\subsubsection{Example 3: Model $\mathcal{M}^{\text{cs(1)}}_B$ [Eq.~(\ref{ModBcomplxsym})]}

\begin{figure}[h]
%\vspace*{-1cm} 
%\centering
\hspace*{-0.1cm}
\scalebox{0.62}{\includegraphics*{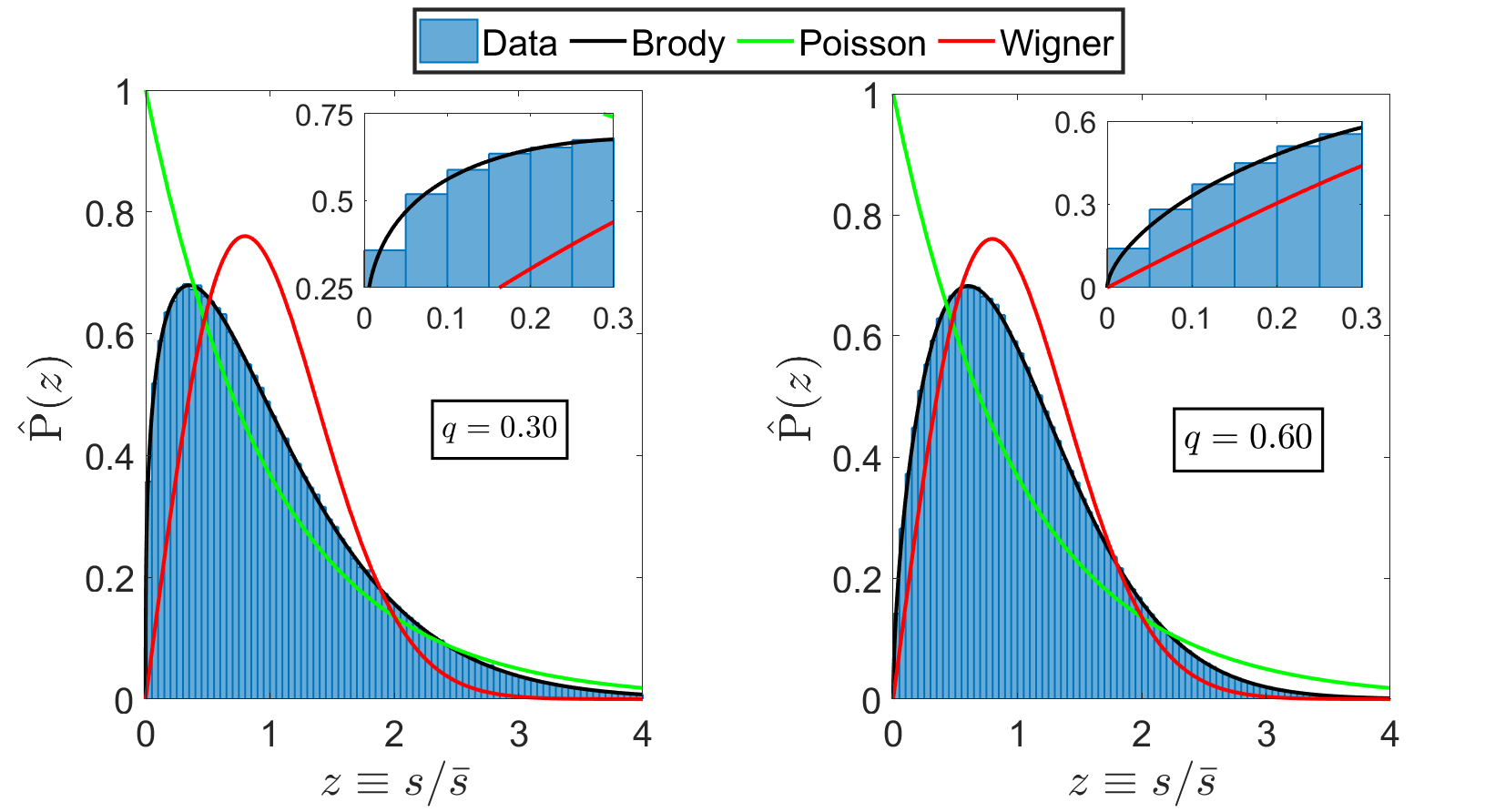}}
%\vspace*{-2.7cm}
\caption{\label{EG3} Eigenvalue spacing distributions for $N=10^6$ realizations of model $\mathcal{M}^{\text{cs(1)}}_B$ [Eq.~(\ref{ModBcomplxsym})]; values of the Brody parameter $q$ are as indicated. Insets show closer views at small spacings.}
\end{figure}

Applying the procedures of Section \ref{procedsEGs} to model $\mathcal{M}^{\text{cs(1)}}_B$ [Eq.~(\ref{ModBcomplxsym})], which is complex and symmetric, yields the results shown in Fig.~\ref{EG3}. The value $\sigma_\text{e}=100$ and $q$-dependent power constants $\{a(q)=-q,b(q)=q+2\}$ were specifically employed in the simulations. Numerics and theory are once again consistent. 

\subsubsection{Example 4: Model $\mathcal{M}^{\text{cs(2)}}_B$ [Eq.~(\ref{ModBcomplxsym2})]}

\begin{figure}[h]
\vspace*{-1.25cm} 
%\centering
\hspace*{-0.29cm}
\scalebox{0.62}{\includegraphics*{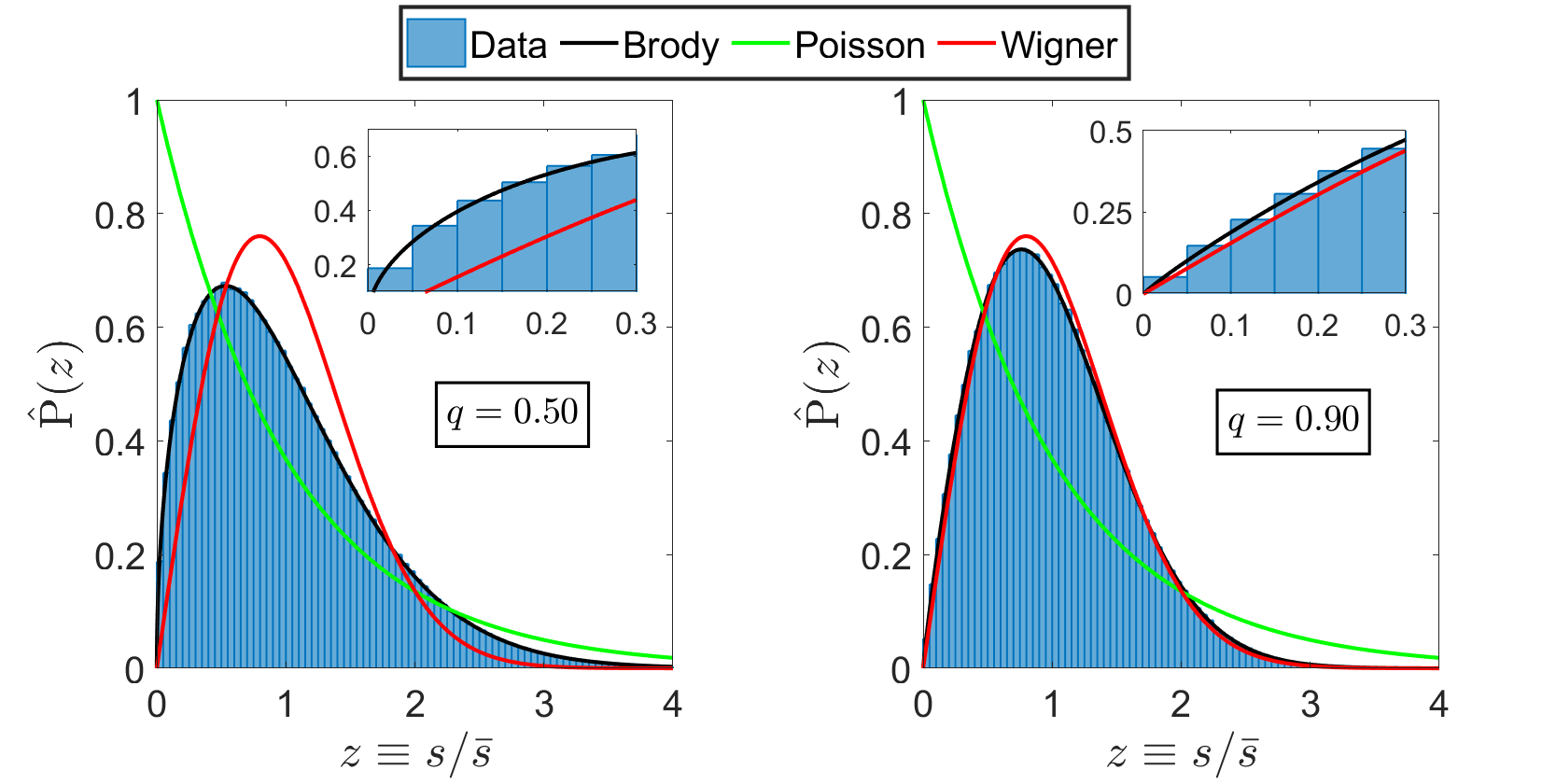}}
%\vspace*{-2.7cm}
\caption{\label{EG4} Eigenvalue spacing distributions for $N=10^6$ realizations of model $\mathcal{M}^{\text{cs(2)}}_B$ [Eq.~(\ref{ModBcomplxsym2})]; values of the Brody parameter $q$ are as indicated. Insets show closer views at small spacings.}
\end{figure}

As an example involving complex-conjugate eigenvalues, results are shown in Fig.~\ref{EG4} for model $\mathcal{M}^{\text{cs(2)}}_B$ [Eq.~(\ref{ModBcomplxsym2})] with $\sigma_\text{e}=1000$ and $q$-dependent power constants $\{a(q),b(q)\}$ given by Eq.~(\ref{QsnsTr0EG}). Note that the random matrices in this case are complex-symmetric. Numerics and theory are again consistent.

\subsubsection{Example 5: Mean Spacing for Model $\mathcal{M}^{(2)}_C$ [Eq.~(\ref{MCeg2})]}

\begin{figure}[h]
\vspace*{-1cm} 
%\centering
%\hspace*{-1cm}
\scalebox{0.41}{\includegraphics*{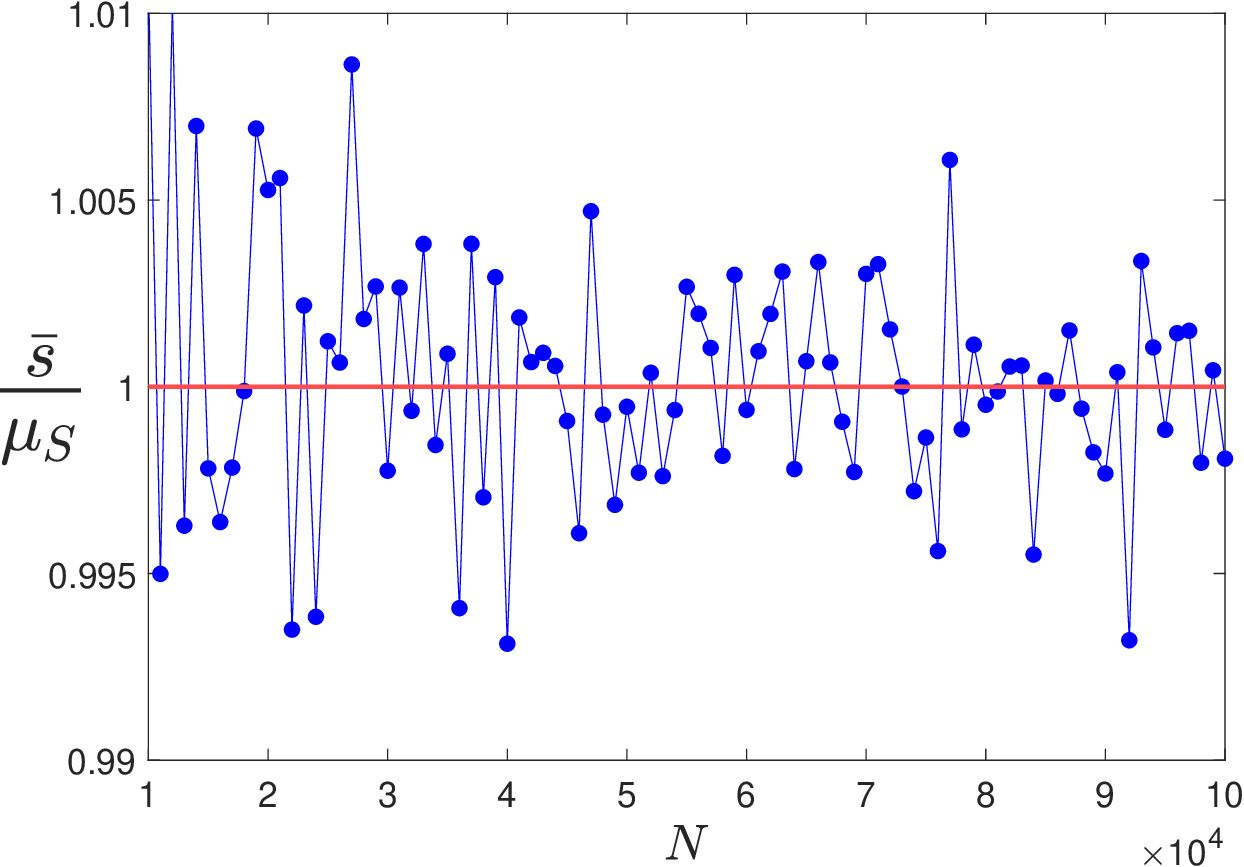}} 
\scalebox{0.41}{\includegraphics*{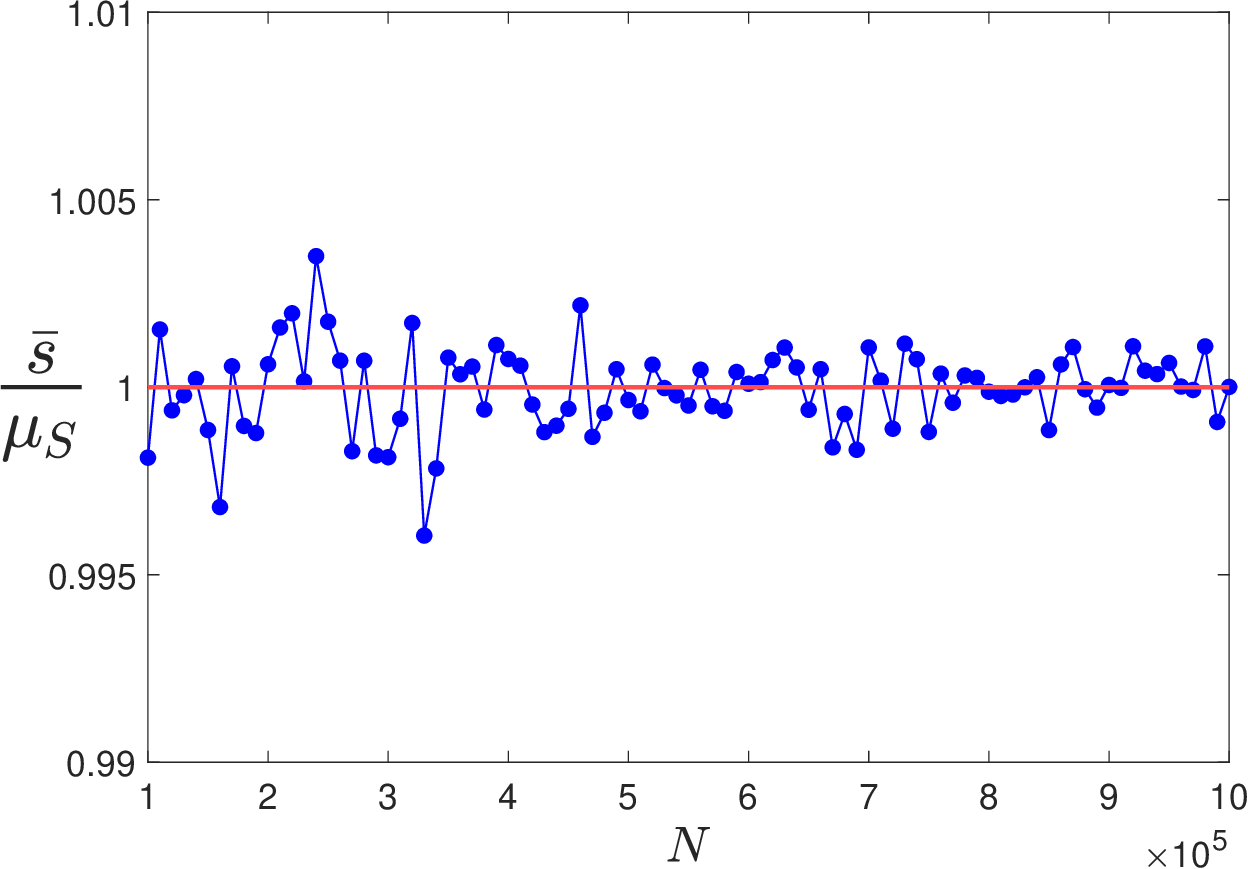}} 
\scalebox{0.42}{\includegraphics*{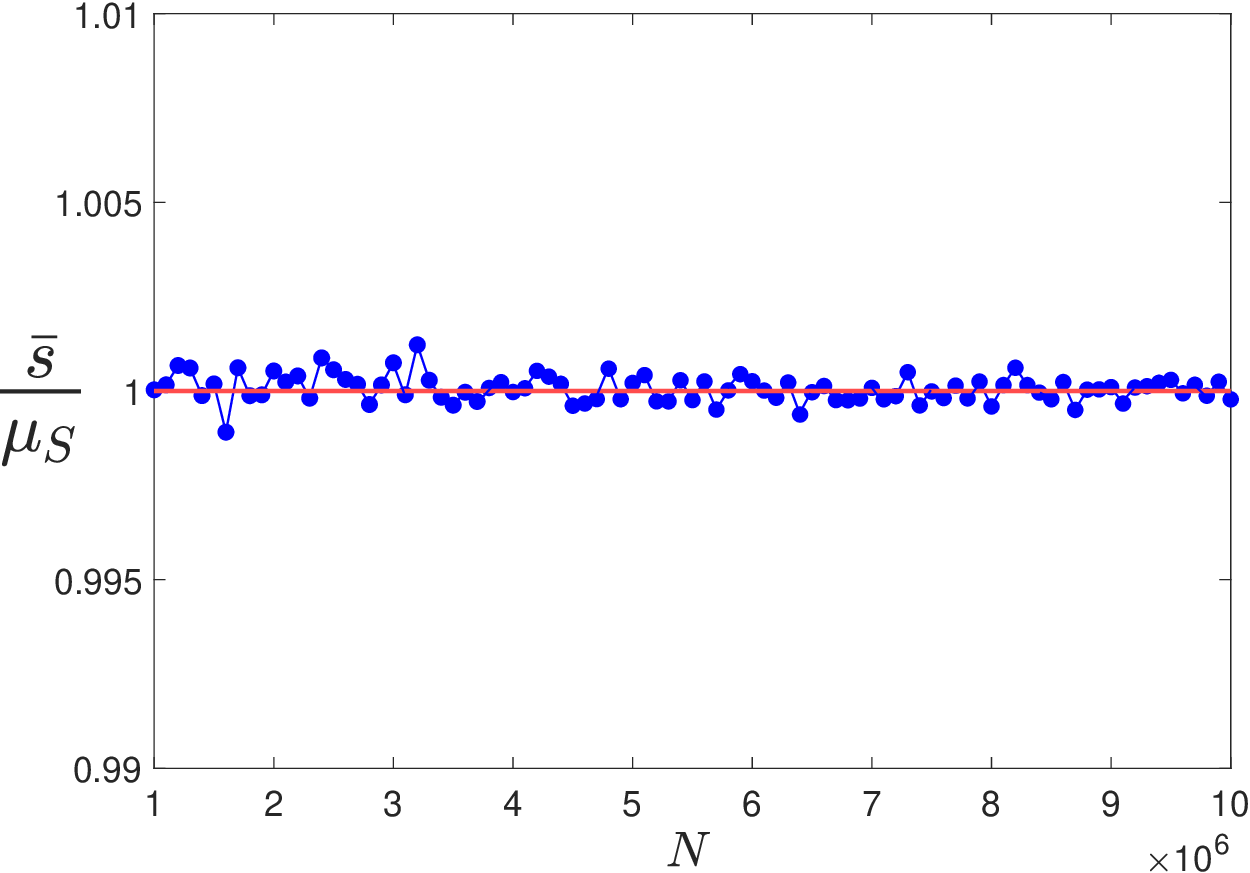}}
%\vspace*{-1.5cm}
\caption{\label{EG5} The ratio $\bar{s}/\mu_S$ at selected values of $N\in\left[10^4,10^7\right]$ for model $\mathcal{M}^{(2)}_C$ [Eq.~(\ref{MCeg2})] with scale parameter $\sigma_\text{e}=\pi$ and Brody parameter $q=1/2$.}
\end{figure}

The law of large numbers (LLN) dictates that the sample mean spacing $\bar{s}\equiv\bar{s}(N)$ converges to the exact (population) mean spacing $\mu_S$ as $N\to\infty$. Since $\mu_S$ is known analytically [c.f., Eq.~(\ref{meanS})], numerical confirmation of the law in any of the studied sub-classses of models is straightforward. As an example, the ratio $\bar{s}/\mu_S$ was calculated in the range $N\in\left[10^4,10^7\right]$ for model $\mathcal{M}^{(2)}_C$ [Eq.~(\ref{MCeg2})] with the results shown in Fig.~\ref{EG5}. Clearly, deviations from unity are smaller when the range of $N$ values is larger, consistent with LLN. 

\subsubsection{Example 6: Cross-Over Transition Between Semi-Poisson and Ginibre Statistics}

\begin{figure}[h]
\vspace*{-1cm} 
%\centering
%\hspace*{-0.23cm}
\scalebox{0.64}{\includegraphics*{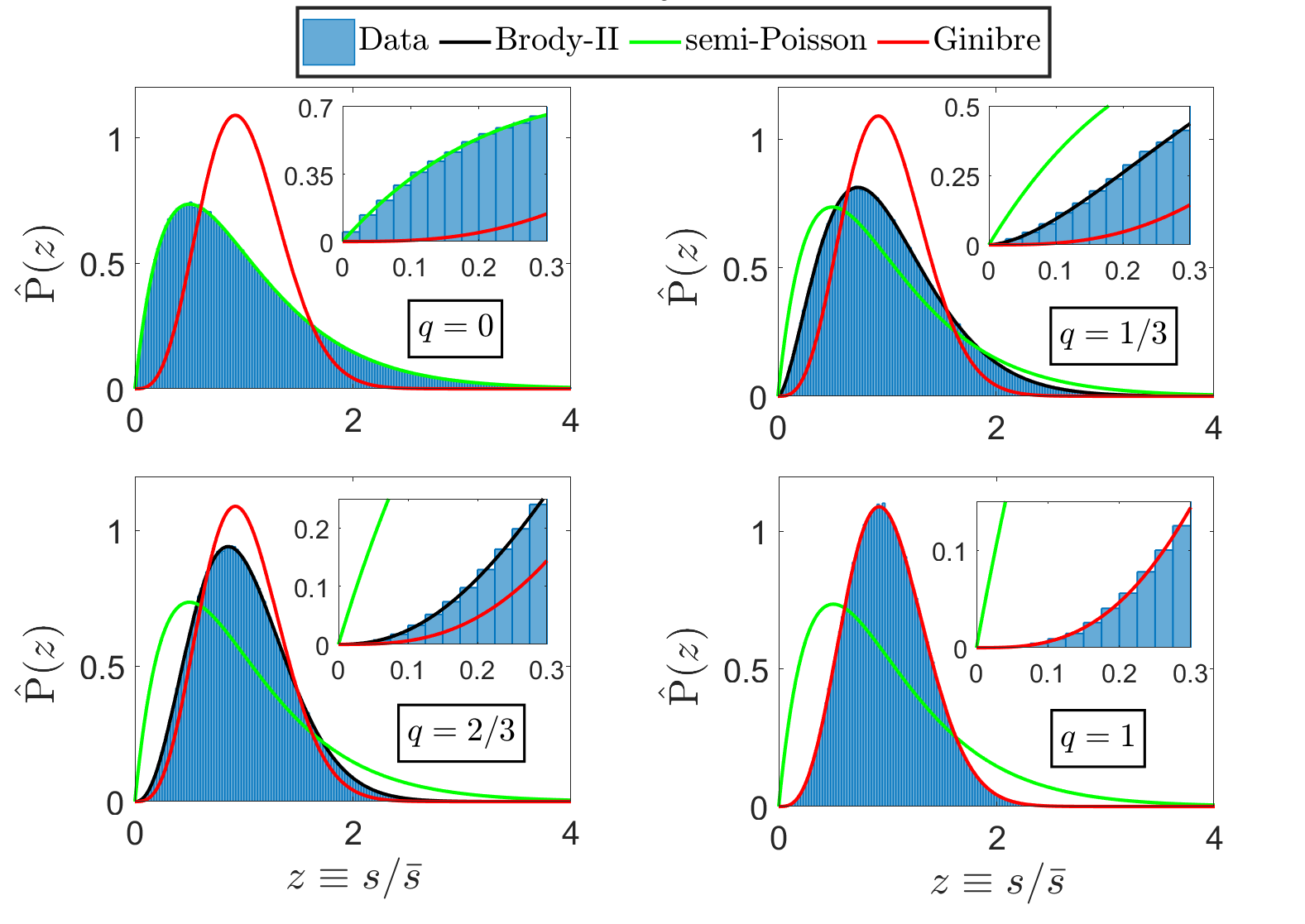}}
%\vspace*{-1.7cm}
\caption{\label{EG6} Eigenvalue spacing distributions for $N=10^6$ realizations of model $\mathcal{M}^{(3)}_C$ [Eq.~(\ref{MCegRV})] with $Y\sim\text{Gamma}(\sigma_\text{g}=10,\xi=2)$, $V\sim\text{Rayleigh}(\sigma_R=5)$, and values of the parameter $q\in[0,1]$ as indicated. The ``Brody-II'' distribution is given by Eq.~(\ref{Eqn:BrodySakhr}). Insets show closer views at small spacings.}
\end{figure}

As an example exhibiting a cross-over transition between semi-Poisson and Ginibre statistics, results are shown in Fig.~\ref{EG6} for model $\mathcal{M}^{(3)}_C$ [Eq.~(\ref{MCegRV})] with $Y\sim\text{Gamma}(\sigma_\text{g}=10,\xi=2)$ and $V\sim\text{Rayleigh}(\sigma_R=5)$. In this example, the sample distributions (i.e., the density histograms) are compared to the theoretical ``Brody-II'' distribution [Eq.~(\ref{Eqn:BrodySakhr})] whose level-repulsion exponent $\beta(q)$ quantifies the fractional degree of level repulsion between the linear (semi-Poisson,$q=0$) and cubic (Ginibre,$q=1$) endpoint cases. Numerics and theory are once again consistent.

\subsubsection{Epilogue: Comment on the Numerical Values of the Scale Parameter $\sigma_\text{e}$}

The numerical value of the scale parameter $\sigma_\text{e}$ can be freely and arbitrarily prescribed. A variety of both small and large values were employed in Examples 1-4; the rationale was merely to demonstrate that the mean-scaled spacing distribution is indeed independent of the numerical value of $\sigma_\text{e}$; the specific values of $\sigma_\text{e}$ employed in the aforementioned examples are not special in any way nor are they contrived values that yield better fits to the data. 
   
\section{Comments}\label{COMMs}

\subsection{Mathematical Comments}

\subsubsection{Comment M1} 

The power-transformed exponential random variables $\left\{Y^{{p_1}\over{q+1}},Y^{{p_2}\over{q+1}},\ldots,Y^{{p_n}\over{q+1}}:n\in\mathbb{N}\right\}$ appearing in the definition of model (\ref{MBG1}) are distinct variables but they are not independent since they are merely different powers (i.e., different functions) of the same random variable $Y$. The matrix elements (which involve different linear combinations of these dependent variables) are therefore generally dependent; it is in this sense that model (\ref{MBG1}) is a ``correlated'' random-matrix model. A secondary but nonetheless fundamental correlation in the model is that the exponents of $Y$ are themselves necessarily related (due to condition (\ref{condhs})). The net effect of this secondary correlation is that the spacing distribution ends up depending on only the distribution of the random variable $Y^{1\over{q+1}}$. The distribution(s) of any other random variable(s) the matrix element(s) could involve (e.g., the variables $\{U,V,X\}$ in model $\mathcal{M}^{(3)}_{A}$ [Eq.~(\ref{MBnsEGRVs})]) do not affect the spacing distribution. 

\subsubsection{Comment M2}
 
If all members of $\mathbb{P}\equiv\left\{{p_1},{p_2},\ldots,{p_n}:n\in\mathbb{N}\right\}$ and $\mathbb{K}\equiv\{k_1,k_2,\ldots,k_{4n+4}:n\in\mathbb{N}\}$ in model (\ref{MBG1}) are constants, then the matrix elements of any cropped or pruned sub-classes of models are (in general) fully correlated in the sense that the values of all matrix elements can be obtained from knowledge of the value of any one matrix element. This perfect correlation is generally destroyed when some members of $\mathbb{P}$ and/or $\mathbb{K}$ are random variables. For example, if $X=0$ in (trimmed) model (\ref{MBnsEGRVs}), then it is clear that the values of the off-diagonal matrix elements cannot be obtained from knowledge of the values of the diagonal matrix elements. Likewise, the value of the lower off-diagonal matrix element in (pruned) model (\ref{MCeg5}) cannot be obtained from sole knowledge of the value of the upper off-diagonal matrix element. Sub-classes of models that are not cropped or pruned are, in general, not perfectly correlated; this is true even when all members of $\mathbb{P}$ and $\mathbb{K}$ are constants.\footnote{The important exception is when the matrix prefactor $\mathcal{A}\neq\text{constant}$, in which case, the matrix elements are, in general, fully correlated.} The takeaway is: condition (\ref{condhs}) does not necessarily imply that the matrix elements are fully correlated. This subtle but important point applies to all $2\times2$ random-matrix models satisfying condition (\ref{condhs}), whether they derive from model (\ref{MBG1}) or not (e.g., class J (\ref{UnrelClass})). 

\subsubsection{Comment M3} 

Beyond model (\ref{MBG1}) introduced in this paper, there are many other models that can be constructed that would satisfy condition (\ref{condhs}). For example, the $2\times2$ random-matrix models 
\begin{subequations}\label{EGDIFF}
\begin{eqnarray}\label{EGDIFF3}
\mathcal{M}_{D1}^{(\pm)}=\left ( \begin{array}{cc}
            \sqrt{Y^{2\over{q+1}}+c^2(q)}  & ~\pm c(q) g(X) \\ [0.15cm]
             \mp \displaystyle{c(q) \over g(X)} & ~-\sqrt{Y^{2\over{q+1}}+c^2(q)}
           \end{array} \right), 
\end{eqnarray}
\begin{eqnarray}\label{EGDIFF1}
\mathcal{M}_{D2}=\left ( \begin{array}{cc}
             X+c(q)i & ~\left(Y^{2\over{q+1}}+c^2(q)\right)^a  \\
             \left(Y^{2\over{q+1}}+c^2(q)\right)^b & ~X-c(q)i
           \end{array} \right), 
\end{eqnarray}
and
\begin{eqnarray}\label{EGDIFF2}
\mathcal{M}_{D3}=\left ( \begin{array}{cc}
             c(q)d_1(X) & ~{1\over2}\sqrt{Y^{2\over{q+1}}-c^2(q)} \\ [0.15cm]
             {1\over2}\sqrt{Y^{2\over{q+1}}-c^2(q)} & ~c(q)d_2(X)
           \end{array} \right), 
\end{eqnarray}
\end{subequations}
do not derive from model (\ref{MBG1}) but nonetheless satisfy condition (\ref{condhs}) and hence have Brody spacing distributions. In models (\ref{EGDIFF}), the random variable $X$ is independent of $Y$ and can have any arbitrary continuous distribution $f_X(x)$, and $c(q)$ is a real $q$-dependent constant that can be arbitrarily prescribed. The other model-specific quantities are as follows: in model (\ref{EGDIFF3}), $g(X)$ is any real function of $X$ having the property $g(X)\neq0$ for all supported values of $X$; in model (\ref{EGDIFF1}), the real exponents $\{a,b\}$ satisfy the condition $a+b=1$ but are otherwise arbitrary; and in model (\ref{EGDIFF2}), the diagonal functions $\{d_1(X),d_2(X)\}$ are any two real functions of $X$ that satisfy $d_1(X)-d_2(X)=1$. In each of models (\ref{EGDIFF3})-(\ref{EGDIFF2}), the off-diagonal matrix elements are independent of the diagonal matrix elements. Note that the off-diagonal elements of $\mathcal{M}_{D3}$ can be real or pure imaginary; the eigenvalues of $\mathcal{M}_{D3}$ are however always real. 

Models (\ref{EGDIFF}) are not real-symmetric nor complex-Hermitian. Hermitian models satisfying condition (\ref{condhs}) can of course also be constructed. One example (that is not a sub-class of model (\ref{MBG1})) is:
\begin{eqnarray}\label{EGDIFF4}
\widetilde{\mathcal{M}}^{(\text{H})}_n(T,X,Y)=\left ( \begin{array}{cc}
            {1\over{D}_n}\left(X^{1\over{q+1}}+Y^{1\over{q+1}}\right)^n & ~c(q)Y^{1\over{q+1}}\exp(iqT) \\ [0.15cm]
             c(q)Y^{1\over{q+1}}\exp(-iqT) & ~{1\over{D}_n}\left(X^{1\over{q+1}}-Y^{1\over{q+1}}\right)^n
           \end{array} \right), 
\end{eqnarray}
where $c$ is an arbitrary (possibly $q$-dependent) real constant, $n$ is a positive integer, and the random variables $X$ and $T$ can have arbitrary continuous distributions $f_X(x>0)$ and $f_T(t)$, respectively (note that $X$ must have strictly positive support). The factor $D_n\equiv{D}_n\left(X^{1\over{q+1}},Y^{1\over{q+1}}\right)$ (which varies with $n$) in the diagonal elements is prescribed such that condition (\ref{condhs}) is satisfied. For example, 
\begin{subequations}\label{EGDIFF4theCs}
\begin{eqnarray}
&&D_1=1, \\
&&D_2=X^{1\over{q+1}}, \\
&&D_3=3X^{2\over{q+1}}+Y^{2\over{q+1}}, \\
&&D_4=X^{1\over{q+1}}\left(X^{2\over{q+1}}+Y^{2\over{q+1}}\right), \\
&&D_5=5X^{4\over{q+1}}+10X^{2\over{q+1}}Y^{2\over{q+1}}+Y^{4\over{q+1}}, \\
&&D_6=X^{1\over{q+1}}\left(3X^{4\over{q+1}}+10X^{2\over{q+1}}Y^{2\over{q+1}}+3Y^{4\over{q+1}}\right).
\end{eqnarray}
\end{subequations}
The $n=1$ case is trivial but is included for completeness. Note that if $T$ is identically zero, then model (\ref{EGDIFF4}) reduces to a real symmetric matrix. 

It is worth mentioning that correlated ``additive'' models satisfying condition (\ref{condhs}) can also be constructed (two simple examples are provided in Appendix \ref{addity}). Incidentally, the generalizations of Cases I and III of sub-class B given in Section \ref{GenCIandIIIModB} are also models that are not directly derivable from model (\ref{MBG1}), but the same comment as above applies to these two cases as well. 
 
\subsubsection{Comment M4} 

Further to Comment M3, a simple way of constructing ensembles of $2\times2$ real-symmetric or complex-Hermitian random matrices that will satisfy condition (\ref{condhs}) is to conjugate the random diagonal matrix 
\begin{eqnarray}\label{RefsD}
\Lambda=\left ( \begin{array}{cc}
             a+{1\over2}W & 0 \\ 
             0 & ~a-{1\over2}W
           \end{array} \right),  
\end{eqnarray}
where $a$ is an arbitrary constant and $W\equiv Y^{1\over{q+1}}$ [c.f., Eq.~(\ref{RayisGG})], by a Haar-random matrix $\mathsf{Q}$ from either $O(2)$ (the group of $2\times2$ orthogonal matrices) or $U(2)$ (the group of $2\times2$ unitary matrices). In either case, the result is a new (``conjugated'') random matrix $\mathsf{A}\equiv\mathsf{Q}\Lambda\mathsf{Q}^{-1}$, which has the same eigenvalues as $\Lambda$ but different (and generally more complicated) entries. The eigenvalue spacing is preserved under conjugation by matrix $\mathsf{Q}$, which means that the spacing distributions for matrices drawn from ensembles $\mathcal{E}_\Lambda$ and $\mathcal{E}_\mathsf{A}$ (the ensemble of Haar-random conjugates of $\Lambda$) will be the same. Since matrices drawn from $\mathcal{E}_\Lambda$ have a Brody spacing distribution then so will matrices drawn from $\mathcal{E}_\mathsf{A}$. 

In the orthogonal case, a Haar-random matrix $\mathsf{Q}_{\text{o}}\in{O(2)}$ is the rotation matrix 
\begin{eqnarray}\label{RotMat}
\mathsf{Q}_{\text{o}}=\left ( \begin{array}{cc}
             \cos\Theta & ~-\sin\Theta \\
             \sin\Theta & \cos\Theta
           \end{array} \right),  
\end{eqnarray} 
where $\Theta\sim\text{Uniform}(0,2\pi)$. By direct calculation, the conjugated random matrix $\mathsf{A}_{\text{o}}$ is:
\begin{eqnarray}\label{RefsConj}
\mathsf{A}_{\text{o}}=\left ( \begin{array}{cc}
            a+{1\over2}W \cos(2\Theta) & {1\over2}W\sin(2\Theta) \\
             {1\over2}W\sin(2\Theta) & ~a-{1\over2}W\cos(2\Theta)
           \end{array} \right), 
\end{eqnarray} 
which has eigenvalue spacing $S=W$. Matrices drawn from the ensemble $\mathcal{E}_{\mathsf{A}_{\text{o}}}$ of conjugated matrices will therefore have a Brody spacing distribution. 

In the unitary case, a general Haar-random matrix $\mathsf{Q}_{\text{u}}\in{U(2)}$ is \cite{Karol}: 
\begin{eqnarray}\label{RotMatU}
\mathsf{Q}_{\text{u}}=\exp(i\Psi)\left ( \begin{array}{cc}
             \exp(i\Phi_1)\cos\Theta & \exp(i\Phi_2)\sin\Theta \\
             -\exp(-i\Phi_2)\sin\Theta & ~\exp(-i\Phi_1)\cos\Theta
           \end{array} \right),  
\end{eqnarray} 
where $\{\Psi,\Phi_1,\Phi_2\}$ are random phases independently and uniformly distributed in the interval $(0,2\pi)$, and $\Theta=\sin^{-1}(\sqrt{\Upsilon})$ with $\Upsilon\sim\text{Uniform}(0,1)$. By direct calculation, the conjugated random matrix $\mathsf{A}_{\text{u}}$ is:
\begin{eqnarray}\label{RefsConjU}
\mathsf{A}_{\text{u}}=\left ( \begin{array}{cc}
            a+{1\over2}W \cos(2\Theta) & ~-{1\over2}W\exp\left[i\left(\Phi_1+\Phi_2\right)\right]\sin(2\Theta) \\
             -{1\over2}W\exp\left[-i\left(\Phi_1+\Phi_2\right)\right]\sin(2\Theta) & a-{1\over2}W\cos(2\Theta)
           \end{array} \right), 
\end{eqnarray} 
whose eigenvalue spacing $S=W$, which again means that the ensemble $\mathcal{E}_{\mathsf{A}_{\text{u}}}$ of conjugated matrices has a Brody spacing distribution. 

While the above results certainly answer the existence question \cite{JohnandLutz} in the traditional setting of Hermitian quantum mechanics, they are limited in scope and have some drawbacks. Model (\ref{RefsConj}) is one specific model and is a rather trivial one in the sense that it can be decomposed as follows:
\begin{eqnarray}\label{RefsConjDecomp}
\mathsf{A}_{\text{o}}=a\left ( \begin{array}{cc}
             1 & 0 \\ 
             0 & 1
           \end{array} \right)+{1\over2}W\left ( \begin{array}{cc}
            \cos(2\Theta) & ~\sin(2\Theta) \\
             \sin(2\Theta) & ~-\cos(2\Theta)
           \end{array} \right)\equiv{a\mathsf{I}}+{1\over2}W\mathsf{F}_{\text{o}},
\end{eqnarray} 
where the matrix $\mathsf{F}_{\text{o}}$ depends only on the random variable $\Theta$ (the diagonal matrix ${a\mathsf{I}}$ does not affect the eigenvalue spacing distribution). The matrix elements of $\mathsf{A}_{\text{o}}$ [Eq.~(\ref{RefsConj})] are also fully correlated. Model (\ref{EGDIFF4}) with $T$ identically zero is, in comparison, non-trivial (for $n\ge3)$ and its matrix elements are not fully correlated. Similar comments apply to model (\ref{RefsConjU}), which can be decomposed as follows:
\begin{eqnarray}\label{RefsConjDecompU}
\mathsf{A}_{\text{u}}&=&a\left ( \begin{array}{cc}
             1 & 0 \\ 
             0 & 1
           \end{array} \right)+{1\over2}W\left ( \begin{array}{cc}
            \cos(2\Theta) & ~-\exp\left[i\left(\Phi_1+\Phi_2\right)\right]\sin(2\Theta) \\
            -\exp\left[-i\left(\Phi_1+\Phi_2\right)\right]\sin(2\Theta) & ~-\cos(2\Theta)
           \end{array} \right) \nonumber \\
           &\equiv&{a\mathsf{I}}+{1\over2}W\mathsf{F}_{\text{u}}(\Phi_1,\Phi_2,\Theta),
\end{eqnarray} 
where the matrix $\mathsf{F}_{\text{u}}$ is independent of $W$. However, unlike the orthogonal case, the matrix elements of $\mathsf{A}_{\text{u}}$ [Eq.~(\ref{RefsConjU})] are not fully correlated.

It should be noted that the diagonal matrix $\Lambda$ need not have the simple form given by Eq.~(\ref{RefsD}); other forms are tenable. In fact, the shortcomings of the above approach to constructing Hermitian ensembles (e.g., the triviality expressed by Eqs.~(\ref{RefsConjDecomp}) and (\ref{RefsConjDecompU})) can sometimes be remedied by employing certain types of diagonal matrices. For example, consider the diagonal matrix:
\begin{subequations}\label{JsAltUeg}
\begin{eqnarray}\label{JsRatD}
\widetilde{\Lambda}=\left ( \begin{array}{cc}
            \displaystyle {W^{a}\over\left(W^{a-1}-W^{b-1}\right)} & 0 \\ 
             0 & \displaystyle~~{W^{b}\over\left(W^{a-1}-W^{b-1}\right)}
           \end{array} \right),~a \neq b  
\end{eqnarray}
whose conjugation by $\mathsf{Q}_{\text{u}}$ [Eq.~(\ref{RotMatU})] yields:
\begin{eqnarray}\label{JsConjMatU}
\widetilde{\mathsf{A}}_{\text{u}}=\left ( \begin{array}{cc}
           \displaystyle {W^{a}\over\left(W^{a-1}-W^{b-1}\right)}-W\sin^2(\Theta) & ~-{1\over2}W\exp\left[i\left(\Phi_1+\Phi_2\right)\right]\sin(2\Theta) \\ [0.10cm]
             -{1\over2}W\exp\left[-i\left(\Phi_1+\Phi_2\right)\right]\sin(2\Theta) & \displaystyle~{W^{b}\over\left(W^{a-1}-W^{b-1}\right)}+W\sin^2(\Theta)
           \end{array} \right),
\end{eqnarray}
which cannot be decomposed in a manner analogous to (\ref{RefsConjDecompU}), that is,
\begin{equation}\label{decompALT}
\widetilde{\mathsf{A}}_{\text{u}}\neq{d(W)}\mathsf{I}+W\widetilde{\mathsf{F}}_{\text{u}}(\Phi_1,\Phi_2,\Theta), 
\end{equation}
\end{subequations}
where $d(W)$ is a real function of the random variable $W$ and the matrix $\widetilde{\mathsf{F}}_{\text{u}}$ is independent of $W$. It is straightforward to verify that the Brody distribution ensues for model (\ref{JsConjMatU}). The choice of diagonal matrix (\ref{JsRatD}) in the unitary case therefore yields a ``non-trivial'' model. Note that the exponents $\{a,b\}$ in model (\ref{JsAltUeg}) are arbitrary (possibly $q$-dependent) real constants (e.g., $\{a=-\sqrt{5},b=q\pi\}$). More generally, these exponents could be real functions of independent random variables (e.g., $\{a(U)=2q\sin(U/3),b(V)=3\cos^2(V)\}$). 

\subsubsection{Comment M5} 

The distributions of the individual matrix elements of a general $2\times2$ random matrix $\mathcal{M}$ are not fundamentally important for determining the eigenvalue spacing distribution of $\mathcal{M}$. According to Eqs.~(\ref{spacsG}) and (\ref{spacsCCF}), the distribution of the mixture of random variables that comprise the quantity $\mathcal{D}(\mathcal{M})$ determines the spacing distribution. Thus, the individual matrix elements need not consist of sums of different power-transformed exponential random variables; it is sufficient that the discriminant $\mathcal{D}(\mathcal{M})$ be proportional to the square of a Weibull random variable with shape parameter $\tau=q+1$:
\begin{equation}\label{gencond}
\mathcal{D}(\mathcal{M})=kL^2,~~k\neq0,~L\sim\text{Weibull}(\kappa=\sigma(\sigma_{1},\sigma_{2},\ldots,\sigma_{l}),\tau=q+1),
\end{equation}
where $\kappa=\sigma(\cdots)$ denotes the fact that the Weibull scale parameter $\kappa$ can depend on any distributional scale parameters $\{\sigma_{1},\sigma_{2},\ldots,\sigma_{l}:l\in\mathbb{N}\}$ the random-matrix model may involve. 

As an illustrative example, consider the following $2\times2$ random-matrix model: 
\begin{eqnarray}\label{EGDIFFCM3}
\mathcal{M}_R=\left ( \begin{array}{cc}
             U^2 & ~{1\over2}R^{A}\left(R^{-{2q\over{q+1}}}+1\right) \\
             {1\over2}R^{B}\left(R^{-{2q\over{q+1}}}-1\right) & -V^2
           \end{array} \right), 
\end{eqnarray}
where $\{U,V\}$ are IID normal random variables with zero mean and variance $\sigma^2_R$, the variable $R\equiv\sqrt{U^2+V^2}$, and the (dependent) random variables $\{A,B\}$ are constrained such that $A+B=4$ but are otherwise arbitrary. Note that the diagonal matrix elements are mutually independent. By direct calculation, 
\begin{equation}
\mathcal{D}\left(\mathcal{M}_R\right)=\left(U^2+V^2\right)^{2\over{q+1}}.
\end{equation}
It can be shown (see Appendix \ref{RayCalc}) that the eigenvalue spacing (c.f., Eq.~(\ref{spacsG}))
\begin{equation}\label{RayisGGRLY}
S=\left(U^2+V^2\right)^{1\over{q+1}}\sim\text{Weibull}\left(\kappa=2\sigma^2_R,\tau=q+1\right), 
\end{equation}
and that the distribution of $Z\equiv {S/\mu_S}$ (i.e., the mean-scaled spacing distribution) is again the Brody distribution. 

Condition (\ref{condhs}) [with $f_Y(y)$ as given by Eq.~(\ref{rayden})] and condition (\ref{gencond}) are not identical; the former is a particular case of the latter. 
There is a fundamental distinction between conditions (\ref{condhs}) and (\ref{gencond}): condition (\ref{condhs}) is a sufficiency condition for model (\ref{MBG1}) to have a Brody spacing distribution, whereas condition (\ref{gencond}) is a more general sufficiency condition that can be applied to evaluate and construct other types of $2\times2$ models. The important general conclusion is that Eq.~(\ref{atBrody}) holds whenever condition (\ref{gencond}) is satisfied regardless of the distributions of the individual matrix elements. Interestingly, this means that many different $2\times2$ random-matrix models will have a Brody spacing distribution. The question of uniqueness has therefore been answered in the negative.

\begin{table}[h]
\vspace*{1cm}
\centering 
\begin{tabular}{ll}
    \hline \hline 
   Model~~~ & ~~~~~Mutually Independent Elements \\ \hline  
   $\mathcal{M}_E^{(2)}$ [Eq.~(\ref{MBsim2})] & ~~~~~diagonal and off-diagonal pairs \\ 
   $\mathcal{M}^{(\text{g})}_{B(I)}$ [Eq.~(\ref{MBGEN1A})] & ~~~~~diagonals \\ 
   $\mathcal{M}^{(\text{g})}_{B(III)}$ [Eq.~(\ref{MBBGEN2A})] & ~~~~~diagonals \\ 
   $\mathcal{M}^{(1)}_G$ [Eq.~(\ref{MCmoreEG})] & ~~~~~off-diagonal and lower-diagonal pairs \\ 
   $\mathcal{M}_C^{(5)}$ [Eq.~(\ref{MCeg5})] & ~~~~~upper diagonal and upper off-diagonal \\
   $\mathcal{M}_{D1}^{(\pm)}$ [Eq.~(\ref{EGDIFF3})] & ~~~~~diagonal and off-diagonal pairs \\ 
   $\mathcal{M}_{D2}$ [Eq.~(\ref{EGDIFF1})] & ~~~~~diagonal and off-diagonal pairs \\ 
   $\mathcal{M}_{D3}$ [Eq.~(\ref{EGDIFF2})] & ~~~~~diagonal and off-diagonal pairs \\ 
   $\mathcal{M}_R$ [Eq.~(\ref{EGDIFFCM3})] & ~~~~~diagonals  
   \\ \hline \hline 
\end{tabular}
\caption{Models having independent matrix elements.}
\label{TabINDs}
\end{table} 

\subsubsection{Comment M6}

As previously mentioned, the full correlation (conditionally) present in pruned and cropped models can (if desired) be selectively destroyed by employing random variables in the summand exponents and/or coefficients (c.f., Comment M2) and/or making use of the functions $\{g_1(X),g_2(X),h(T)\}$ in the general model (\ref{MBG1}) (c.f., the paragraph preceding Section \ref{subbyA}). For reference, Table \ref{TabINDs} lists the models given in the current paper that possess independent matrix elements. In general, mutual independence between any pair of matrix elements can be effected by judicious placement of random functions. For instance, it is clear from inspection of Table \ref{TabINDs} that no example was given that possessed mutually independent off-diagonal elements. This particular independence structure can be easily effected in many ways using random functions. A simple example is:
\begin{eqnarray}\label{MBLP1}
   \left ( \begin{array}{cc}
             Y^{1\over{q+1}} & {1\over2}\sin^2(2qV) \\
             {1\over2}Y^{2\over{q+1}} & ~2\sin^2(qV)Y^{1\over{q+1}}
           \end{array} \right),
\end{eqnarray}
where the random variable $V$ is independent of $Y$ and can have any arbitrary continuous distribution $f_V(v)$. The above example derives from applying the generalization discussed in Section \ref{GenConsts} to sub-class A, more specifically, allowing some of the sub-class A coefficients $\{c_j\}$ to be random variables as follows: $\{c_1,c_2,c_3,c_4\}=\left\{1,{1\over2}\sin^2(2qV),{1\over2},2\sin^2(qV)\right\}$. 

\subsection{Physics Comments}

\subsubsection{Comment P1} 

In defining model (\ref{MBG1}), no specific conditions were imposed that demanded the matrices be real and/or symmetric. The interesting consequence is that complex non-Hermitian $2\times2$ random matrices can have spacing statistics intermediate between Poisson and Wigner. The real-symmetric structure of $2\times2$ GOE matrices is not required. GOE matrices owe their real and symmetric structure to assumptions and conditions that originate from quantum mechanics. The fact that GOE matrices are symmetric, for example, originally stems from the general assumption that quantum Hamiltonians are Hermitian. This assumption is useful in quantum mechanics since it ensures that all eigenvalues are real, but it is not required. In other words, hermiticity is a sufficient but not necessary condition for the eigenvalues to be real. In sub-class A (for example), the eigenvalues are guaranteed to be real by imposition of condition (\ref{MB2}) when the model constants are real or conditions (\ref{condsModAcmplx}) when the constants are complex; hermiticity is not required. Hermiticity is not fundamentally important in the present context due to the small size of the matrices which allows for other tractable (and less restrictive) conditions that can be imposed to ensure the eigenvalues are real; when dealing with large matrices, on the other hand, hermiticity is more imperative. In summary, model (\ref{MBG1}) has no built-in quantum-mechanical assumptions or conditions.

\subsubsection{Comment P2} 

The $q=0$ limit of model (\ref{MBG1}) can (subject to condition (\ref{condhs})) serve as a new class of $2\times2$ (correlated) random-matrix models having Poisson spacing statistics. Valid sub-classes of models generated from the $q=0$ limit of model (\ref{MBG1}) are interesting in the respect that they illustrate that the matrix elements need not be IID nor do any elements need to be Poisson (exponentially) distributed in order for the eigenvalues to have Poissonian spacings. For example, the eigenvalue spacings of real non-symmetric random matrices of the (pruned) form (c.f., Section \ref{SignifB})
\begin{eqnarray}\label{PssnEG}
   \left ( \begin{array}{cc}
             c_1Y^{1\over3}+c_2 & c_3Y^{2\over3}+c_4 \\
             c_5Y^{2\over3} +c_6 & c_7Y+c_8
           \end{array} \right)
\end{eqnarray} 
will have a Poisson distribution as long as the real constants $\{c_j:j=1,\ldots,8\}$ are such that condition (\ref{condhs}) is satisfied.\footnote{Sufficient conditions on $\{c_j\}$ are: (i) $c_2=c_8$; (ii) $c_4c_6=0$; (iii) $4c_3c_5-2c_1c_7=0$; (iv) $c^2_1+4(c_4c_5+c_3c_6)=0$. One solution (from an infinite number) is: $\{c_1=-2,~c_3=c_5=1,~c_2=c_4=c_8=0,~c_6=c_7=-1\}$.} Note that none of the matrix elements of model (\ref{PssnEG}) are exponentially distributed (except the lower diagonal element when $c_7>0$ and $c_8=0$). In general, the matrices need not even be real or Hermitian as the $q=0$ limits of models (\ref{MBegcmplx}) and (\ref{ModBcomplxsym}) illustrate. Complex eigenvalues can as well have Poisson spacing statistics (c.f., Section \ref{CCevsSec}). The interesting point is that (real or complex) random matrices whose elements are fully correlated can generate (real or complex) eigenvalues whose spacings are completely uncorrelated. Non-trivial $2\times2$ random-matrix models having Poisson spacing statistics are curiously absent in the literature.\footnote{``Non-trivial'' in this context means $2\times2$ random-matrix models where all matrix elements are non-zero, independent, and not exponentially distributed.} Interestingly, there does however exist a simple $N\times N$ diagonal model whose eigenvalue spacing distribution converges to Eq.~(\ref{Eqn:Poisson}) as $N\to\infty$ (see Ref.~\cite{Magy} and references therein). 

Although not of primary interest in the current paper, it should be noted that the $q=0$ limit of model (\ref{MBG1}) with $f_Y(y)$ given by (\ref{gammaden}) (rather than (\ref{rayden})) can generate novel classes of real or complex $2\times2$ random matrices possessing real or complex eigenvalues with semi-Poissonian spacings (hermiticity is again not a requirement when the matrices are complex).

\subsubsection{Comment P3} 

The $q=1$ limit of model (\ref{MBG1}) can (subject to condition (\ref{condhs})) serve as a new class of $2\times2$ (correlated) random-matrix models having Wigner spacing statistics. The existence of this class of models reinforces the important point that the independence and quantum-mechanical assumptions associated with the construction of the GOE are sufficient but not necessary conditions for the spacing distributions of $2\times2$ random matrices to be Wignerian. More explicitly, the elements of a $2\times2$ random matrix need not be IID nor does the matrix need to be real and/or symmetric in order for its eigenvalues to have a Wigner spacing distribution. As a simple but concrete example, the (possibly complex) sub-class of matrices (c.f., Eq.~(\ref{MBnsEG}))
\begin{eqnarray}\label{MBnsEGspec2}
   \mathcal{M}^{(1)}_A(q=1)=\left ( \begin{array}{cc}
             X+c_1\sqrt{Y} & c_2{1\over\sqrt{Y}} \\
             c_3\sqrt{Y^3} & X+c_4\sqrt{Y}
           \end{array} \right)
\end{eqnarray}
will (under the condition that $k_A\equiv(c_1-c_4)^2+4c_2c_3$ is real and non-zero) have Wigner spacing distributions. A separate and physically important example that illustrates this point can be found in Ref.~\cite{Ahmed2003} (see also Ref.~\cite{Robnik2007} for an example of a real non-symmetric $2\times2$ random-matrix model having a Wigner spacing distribution).\footnote{Note that, in Refs.~\cite{Ahmed2003,Robnik2007}, the matrix elements are Gaussian distributed. An important point that is often glossed over in the literature is that when the classical independence assumption is relaxed (as is the case in model (\ref{MBG1})), the matrix elements need not be Gaussian distributed nor are they required to be in order for the eigenvalue spacings to be Wigner distributed. The preceding is true even when independence is stipulated provided the matrix elements are allowed to be non-identically distributed.}

Analogous comments apply to the $q=1$ limit of model (\ref{MBG1}) with $f_Y(y)$ given by (\ref{gammaden}) (rather than (\ref{rayden})) which can generate novel classes of real or complex $2\times2$ random matrices possessing real or complex eigenvalues whose spacings follow the Ginibre distribution (\ref{Ginibre}). This is interesting, both mathematically and physically, since it clearly shows that the elements of a $2\times2$ random matrix need not be IID and/or Gaussian-distributed nor does the matrix need to be complex in order for its eigenvalues (which also need not be complex) to have a Ginibre spacing distribution (and hence cubic level repulsion). Moreover, the presence or absence of hermiticity/symmetry is irrelevant (at least when correlations between matrix elements are permitted). The preceding are fundamentally important points that (to the author's knowledge) have not been discussed in the literature. 

\subsubsection{Comment P4} 

The real symmetric models in Section \ref{RSegsSCs} may seem unphysical due to the fully-correlated nature of the matrix elements. Random matrix models having highly-correlated entries are actually not a novelty in physics (see, for example, Refs.~\cite{CRP1,CRP2,Alt2015,CRP3,CRP4} and references contained therein). The success of classical RMT (in physics and other fields) does not rely on all of its specific model assumptions. As mentioned in the Introduction, there is a large body of mathematical research that rigorously shows that the classical results of RMT do not depend on the specific nature of the matrix elements (including whether or not the entries are correlated) but rather only on the symmetry class of the matrix (e.g., real symmetric, complex Hermitian, etc.). This is the celebrated spectral universality that was originally conjectured by Wigner, Dyson, Gaudin, and Mehta \cite{Mehta}. The classical independence assumption used in the construction of the GOE is an assumption that simplifies the mathematical analysis and allows for analytical results but has no direct connection to quantum mechanics since the matrix elements of quantum Hamiltonian systems are typically not independent (see, for example, Refs.~\cite{nonindR1,nonindR2}). The matter of whether the matrix elements are uncorrelated, partially correlated, or fully correlated, is therefore not a general concern. 

\subsubsection{Comment P5} 

Although model (\ref{MBG1}) can generate real-symmetric and complex-Hermitian sub-classes of models, it is most suitable for generating real non-symmetric and complex non-Hermitian models. In terms of application, model (\ref{MBG1}) with arbitrary $f_Y(y)$ and bounded $q\ge0$, could (broadly speaking) serve as a general one-parameter random-matrix model for two-state non-Hermitian (classical or quantum) Hamiltonian systems. Many interesting physical examples of such two-state systems can be found in the literature (see, for example, Refs.~\cite{WRSG,PTG1,PTG2,NHP} and references therein). Model (\ref{MBG1}) could therefore find application in many different non-Hermitian contexts and settings where the two-state approximation is useful. Specific examples could include (to name only two): the statistics of avoided resonance crossings in open chaotic billiards (e.g., \cite{Poli}) and the eigenvalue spacing statistics of dielectric microcavities with different refractive indices (e.g., \cite{koreans}). 

\section{Conclusion}\label{CONCLSN}

A class of $2\times2$ random-matrix models was introduced for which the Brody distribution is the exact eigenvalue spacing distribution. To the author's knowledge, this is the first class of random matrices that has been found to possess this specific statistical property. Unlike GOE matrices, the matrix elements of this class of random matrices are not IID nor are they Gaussian-distributed. Apart from random offset functions, the matrix elements are constrained sums of an exponential random variable raised to different (and possibly variable) fractional powers involving the Brody parameter. The numerators of these fractional exponents are subject only to condition (\ref{condhs}) and can otherwise be arbitrarily prescribed. When all of the numerators are constants, some summands could be Weibull random variables with various scale and shape parameters (involving the Brody parameter)\footnote{Cautionary note: linear combinations of Weibull random variables are not Weibull-distributed.}; this will not be the case when the numerator of an exponent is negative (e.g., the upper off-diagonal matrix element of model $\mathcal{M}^{(1)}_{A}$ [Eq.~(\ref{MBnsEG})]) or when multiplicative constants are negative (e.g., the upper diagonal matrix element of model $\mathcal{M}^{(1)}_{C}$ [Eq.~(\ref{MCeg1})]). 

Real matrices that derive from model (\ref{MBG1}) are not necessarily symmetric. The quantum hermiticity condition, which ensures that all eigenvalues are real, was not imposed in defining model (\ref{MBG1}) and is (in the present $2\times2$ context) a sufficient but not necessary condition for the eigenvalues to be real. As Examples 2-4 in Section \ref{NumEgs} explicitly demonstrate, complex non-Hermitian $2\times2$ random matrices whose eigenvalues are real (or complex) can possess spacing statistics intermediate between Poisson and Wigner. The author is not aware of any other complex non-Hermitian random-matrix models (or desymmetrized non-Hermitian physical systems) whose real or complex eigenvalues possess such statistics. It is important to emphasize that the classical assumption of IID matrix elements has been dropped in the present model and the question of whether the preceding statement about non-Hermitian matrices holds when the IID assumption is not removed remains open. There are also questions related to pseudo-symmetry \cite{KA2017} and/or pseudo-hermiticity \cite{FR2021} that are pertinent and interesting but such questions are beyond the scope of the present paper. 

Finally, as discussed in Comment M5, the individual matrix elements need not be composed of power-transformed exponential random variables. Thus, there is some flexibility in assigning the distributions of the matrix elements. The immediate question then is whether other models having more tunable degrees of correlation (i.e., more flexibility with the assignment of independence between matrix elements) can be constructed. This is indeed possible. The author hopes to introduce a simple but versatile class of semi-correlated $2\times2$ random-matrix models in a sequel paper.
     
%\begin{acknowledgments}
%TBA
%\end{acknowledgments}

\appendix

\section{Statistical Distributions}\label{StatsStuff}

The following can be found in many statistics textbooks (see, for example, Ref.~\cite{Norman}).

\subsection{The Weibull Distribution}\label{AppWBL}

Let the random variable $W\sim\text{Weibull}(\kappa,\tau)$, that is, let $W$ have a Weibull distribution with scale parameter $\kappa>0$ and shape parameter $\tau>0$. The p.d.f. of $W$ is then given by:
\begin{equation}\label{GGD}
f_W(w;\kappa,\tau)={\tau\over\kappa} w^{\tau-1}\exp\left(-{w^\tau\over\kappa}\right),~w>0.
\end{equation}
The mean of $W$ is:
\begin{equation}\label{meanW}
\mu_W(\kappa,\tau)=\left({1\over\kappa}\right)^{-{1/\tau}}\Gamma\left(1+{1\over\tau}\right).
\end{equation}

\subsection{Generalized Gamma Distribution}\label{AppGG}

Let an arbitrary random variable $V\sim\text{GG}(\ell,\omega,\Omega)$, that is, let $V$ have a generalized gamma distribution with scale parameter $\ell>0$ and shape parameters $\omega>0$ and $\Omega>0$. The p.d.f. of $V$ is then given by:
\begin{equation}\label{GGD2}
f_V(v;\ell,\omega,\Omega)={\Omega\over\ell^\omega\Gamma(\omega/\Omega)} v^{\omega-1}\exp\left[-\left(v\over\ell\right)^\Omega\right],~v>0.
\end{equation}
The mean of $V$ is:
\begin{equation}\label{meanW2}
\mu_V(\ell,\omega,\Omega)=\ell{\Gamma\left({(\omega+1)\over\Omega}\right)\over\Gamma\left({\omega\over\Omega}\right)}.
\end{equation}
Let $m>0$ and $p>0$ be arbitrary real constants. Then
\begin{equation}\label{GGP1}
mV\sim\text{GG}(m\ell,\omega,\Omega),
\end{equation}
\begin{equation}\label{GGP2}
V^{p_{}}\sim\text{GG}\left(\ell^{p},{\omega\over p},{\Omega\over p}\right). 
\end{equation}

\subsection{Gamma Distribution}\label{AppGamma}

In the special case $\Omega=1$, the generalized gamma distribution reduces to the gamma distribution with scale parameter $\ell\equiv\sigma_\text{g}>0$ and shape parameter $\omega\equiv\xi>0$:
\begin{equation}\label{GGD3}
f_V(v;\sigma_\text{g},\xi)={1\over\sigma_\text{g}^\xi\Gamma(\xi)} v^{\xi-1}\exp\left(-{v\over\sigma_\text{g}}\right),~v>0.
\end{equation}
If an arbitrary random variable $X\sim\text{Gamma}(\sigma_\text{g},\xi)$ and $p>0$, then 
\begin{equation}\label{GtoGG}
X^{p_{}}\sim\text{GG}\left(\ell=(\sigma_\text{g})^{p},\omega=\xi/p,\Omega=1/p\right).
\end{equation}

\section{Derivation of Distribution (\ref{Eqn:BrodySakhr})}\label{sPtoGinPf}

The derivation of distribution (\ref{Eqn:BrodySakhr}) closely parallels the derivation of the Brody distribution given in Section \ref{NNSD}. The task in this appendix is to determine the distribution of the random spacing $S$ as given by Eq.~(\ref{spacs}) when $Y\sim\text{Gamma}(\sigma_\text{g},\xi=2)$ (instead of $Y\sim\text{Exp}(\sigma_{\text{e}})$). This can be efficiently accomplished by exploiting property (\ref{GtoGG}), which immediately yields: 
\begin{equation}\label{VisGG}
V \equiv Y^{1\over{q+1}}\sim\text{GG}\left(\ell=(\sigma_\text{g})^{1\over{q+1}},\omega=2(q+1),\Omega=q+1\right),
\end{equation}
which (according to definition (\ref{GGD2}) of the GG distribution) yields the following p.d.f.~of $V$:
\begin{equation}\label{prepreB1S}
f_V(v;\sigma_\text{g},q) = {(q+1)\over\sigma^2_\text{g}} v^{2q+1} \exp\left(-{v^{q+1}\over\sigma_\text{g}}\right).
\end{equation} 
Transforming variables, the p.d.f.~of $S$ [c.f., Eq.~(\ref{spacs})] is thus:
\begin{eqnarray}\label{preB1S}
f_S(s)&=&f_{\sqrt{k}V}(s)={1\over \sqrt{k}}f_V\left(v={s\over \sqrt{k}}\right) \nonumber \\
&=&{{(q+1)}\over{\sqrt{k}\sigma^2_\text{g}}}\left(s\over\sqrt{k}\right)^{2q+1}\exp\left[-{1\over\sigma_\text{g}}\left(s\over\sqrt{k}\right)^{q+1}\right],
\end{eqnarray}
where $k>0$ is the discriminant constant found in condition (\ref{condhs}). Note that parametric dependences have been suppressed (for notational brevity) in the first line of Eq.~(\ref{preB1S}).

The distribution of the standardized or mean-scaled eigenvalue spacing $Z\equiv {S/\mu_S}$ can then be calculated as in Section \ref{NNSD} using the transformation equation (\ref{toBrody}), where in the present case, 
\begin{equation}\label{meanSsPGin}
\mu_S=\mu_{\sqrt{k}V}=\sqrt{k}\mu_V=\sqrt{k}\sigma_\text{g}^{1\over{q+1}}\Gamma\left({{2q+3}\over{q+1}}\right),
\end{equation} 
where the last equality follows directly from (\ref{meanW2}) with parameter values as specified in (\ref{VisGG}). Performing the necessary algebra then yields distribution (\ref{Eqn:BrodySakhr}). Note that the unscaled spacing distribution (\ref{preB1S}) depends on both the scale parameter $\sigma_\text{g}$ in Eq.~(\ref{gammaden}) and the discriminant constant $k$, whereas the mean-scaled spacing distribution (\ref{Eqn:BrodySakhr}) has no such dependencies.

\section{Additive-Type Correlated Models: Two Simple Examples}\label{addity}

Some authors have characterized the Poisson to GOE transition using so-called additive random-matrix models (see, for example, Refs.~\cite{LH,Wett12}). In this short appendix, two simple examples of (correlated) additive random-matrix models satisfying condition (\ref{condhs}) are provided as an addendum to Comment M3. Their purpose is purely illustrative and they represent the simplest possible scenario. In the present context, the simplest type of additive random-matrix model is of the form:
\begin{equation}
\mathcal{M}_+(Y;q)=\mathcal{M}_{P}(Y)+q\mathcal{M}_{Q}(Y;q),
\end{equation}
where the entries of $\mathcal{M}_{P}(Y)$ do not depend on the parameter $q$. There is no stipulation on the eigenvalue spacing statistics of $\mathcal{M}_{Q}(Y;q)$, but $\mathcal{M}_{P}(Y)$ should be of an appropriate form such that it has Poisson spacing statistics. 

Let 
\begin{eqnarray}\label{Thetafncn}
\Upsilon(q)=\left \{ \begin{array}{cr}
             0 & ~~\text{if}~q=0 \\
             {1\over{q}} & ~~\text{if}~q\neq0
           \end{array} \right..
\end{eqnarray}
The two examples are then as follows.

\noindent \textbf{Example 1}:
\begin{eqnarray}\label{MAM1}
\mathcal{M}^{(1)}_+(Y;q)&=& \left( \begin{array}{cc}
             0 & ~{q\over2}\left(Y^{-{2q\over{q+1}}}-\Upsilon(q)\right) \\
             {q\over2}Y^2\Upsilon(q) & Y
           \end{array} \right) \nonumber \\
&=&\left( \begin{array}{cc}
             0 & -{1\over2} \\
             0 & Y
           \end{array} \right) + q \left ( \begin{array}{cc}
             0 & ~{1\over2}Y^{-{2q\over{q+1}}} \\
             {1\over2}Y^2\Upsilon(q) & 0
           \end{array} \right) \nonumber \\ 
&\equiv&\mathcal{M}^{(1)}_{P}(Y)+q\mathcal{M}^{(1)}_{Q}(Y;q).
\end{eqnarray}
By direct calculation:
\begin{eqnarray}
\mathcal{D}\left(\mathcal{M}_+^{(1)}\right)=\left \{ \begin{array}{cr}
             Y^2 & ~~\text{if}~q=0 \\
             qY^{2\over{q+1}} & ~~\text{if}~q\neq0
           \end{array} \right..
\end{eqnarray}

\noindent \textbf{Example 2}:
\begin{eqnarray}\label{MAM2}
\mathcal{M}^{(2)}_+(Y;q)&=& \left( \begin{array}{cc}
             Y & ~qY^a\left(Y^{-{q\over{q+1}}}+\Upsilon(q)\right) \\
             qY^{b}\left(Y^{-{q\over{q+1}}}-\Upsilon(q)\right) & -Y
           \end{array} \right) \nonumber \\
&=&\left( \begin{array}{cc}
             Y & 0 \\
             -Y^{b} & -Y
           \end{array} \right) + q \left ( \begin{array}{cc}
             0 & ~Y^a\left(Y^{-{q\over{q+1}}}+\Upsilon(q)\right) \\
             Y^{b-{q\over{q+1}}} & 0
           \end{array} \right) \nonumber \\ 
&\equiv&\mathcal{M}^{(2)}_{P}(Y)+q\mathcal{M}^{(2)}_{Q}(Y;q),
\end{eqnarray}
where the constants $\{a,b\}$ are subject to the condition that $a+b=2$ but are otherwise arbitrary. By direct calculation:
\begin{eqnarray}
\mathcal{D}\left(\mathcal{M}_+^{(2)}\right)=\left \{ \begin{array}{cr}
             4Y^2 & ~~\text{if}~q=0 \\
             4q^2Y^{2\over{q+1}} & ~~\text{if}~q\neq0
           \end{array} \right..
\end{eqnarray}
Thus, in each case, Poisson statistics ensue when $q=0$ and Brody statistics when $q\neq0$. 

\section{Derivation of Result (\ref{RayisGGRLY})}\label{RayCalc}

The derivation of (\ref{RayisGGRLY}) begins with recognition of the fact that $R\equiv\sqrt{U^2+V^2}$ is actually a Rayleigh random variable with scale parameter $\sigma_R>0$ (i.e., $R\sim\text{Rayleigh}(\sigma_R)$). The p.d.f. of $R$ is thus: 
\begin{equation}\label{raydenLKRLLY}
f_R(r;\sigma_R)={1\over\sigma^2_R}r\exp\left(-r^2/2\sigma^2_R\right),~r>0.
\end{equation}
Note that $R$ has a Wigner distribution when $\sigma_R=\sqrt{2/\pi}$. Suppose then that the random eigenvalue spacing $S$ is such that 
\begin{equation}\label{RayisGGRLLY}
S=kR^{2\over{q+1}},~k>0.
\end{equation}
To determine the distribution of $S$, we first observe that the Rayleigh distribution (\ref{raydenLKRLLY}) is a special case of the generalized gamma (GG) distribution (see Appendix \ref{AppGG} for pertinent details on the GG distribution). In particular,
\begin{equation}\label{RayisGGapp}
R\sim\text{GG}\left(\sqrt{2}\sigma_R,2,2\right).
\end{equation}
Using properties (\ref{GGP1}) and (\ref{GGP2}) of the GG distribution with $m=k$ and $p=2/(q+1)$ then immediately yields:
\begin{equation}\label{SisGG}
S\sim\text{GG}\left(k(2\sigma^2_R)^{1\over{q+1}},q+1,q+1\right).
\end{equation} 
Using (\ref{GGD2}) with parameters $\ell=k(2\sigma^2_R)^{1\over{q+1}}$ and $\omega=\Omega=q+1$, the p.d.f.~of $S$ is thus:
\begin{equation}\label{preBrody2}
f_S(s)={{(q+1)}\over{(2\sigma^2_R)k}}\left(s\over k\right)^q\exp\left[-\left({1\over2\sigma^2_R}\right)\left(s\over k\right)^{q+1}\right].
\end{equation}
Result (\ref{RayisGGRLY}) immediately follows upon comparing distributions (\ref{preBrody2}) and (\ref{GGD}) in the special case when $k=1$. 

The distribution of the standardized or mean-scaled spacing $Z\equiv {S/\mu_S}$ can then be calculated as before using Eq.~(\ref{toBrody}), where in the present case, 
\begin{equation}\label{meanS2}
\mu_S=k(2\sigma^2_R)^{1\over{q+1}}\Gamma\left({{q+2}\over{q+1}}\right),
\end{equation} 
which follows directly from (\ref{meanW2}). Performing the necessary algebra again yields the Brody distribution [Eq.~(\ref{Eqn:Brody})].


\begin{thebibliography}{99} 

\bibitem{classi} D. Wintgen and H. Friedrich, Phys. Rev. A \textbf{35}, 1464(R) (1987). 

\bibitem{Haake91} F. Haake \textit{et al.}, Phys. Rev. A \textbf{44}, R6161 (1991).

\bibitem{Graf92} H.-D. Graf \textit{et al.}, Phys. Rev. Lett. \textbf{69}, 1296 (1992). 

\bibitem{Kudrolli94} A. Kudrolli, S. Sridhar, A. Pandey, and R. Ramaswamy, Phys. Rev. E \textbf{49}, R11 (1994).

\bibitem{Leitner97} D.M. Leitner, Phys. Rev E \textbf{56}, 4890 (1997).

\bibitem{Cheng02} N.-P. Cheng, Z.-Q. Chen, and H. Chen, Chin. Phys. Lett. \textbf{19}, 309 (2002).

\bibitem{Petit15} J. Mur-Petit and R. A. Molina, Phys. Rev E \textbf{92}, 042906 (2015).

\bibitem{Roy17} K. Roy \textit{et al.}, Europhys. Lett. \textbf{118}, 46003 (2017).

\bibitem{Zhang19} R. Zhang \textit{et al.}, Chin. Phys. B \textbf{28}, 100502 (2019).

\bibitem{Sierant19} P. Sierant and J. Zakrzewski, Phys. Rev. B \textbf{99}, 104205 (2019).

\bibitem{Mehta} M.L. Mehta, \textit{Random Matrices}, 3rd Ed. (Elsevier, San Diego, 2004).

\bibitem{Haake} F. Haake, \textit{Quantum Signatures of Chaos}, 3rd ed. (Springer, Berlin, 2010).

\bibitem{Brody} T.A. Brody, Lett. Nuovo Cimento \textbf{7}, 482 (1973). 

\bibitem{BR} M.V. Berry and M. Robnik, J. Phys. A \textbf{17}, 2413 (1984). 

\bibitem{CGR} E. Caurier, B. Grammaticos, and A. Ramani, J. Phys. A \textbf{23}, 4903 (1990).

\bibitem{Izzy} F.M. Izrailev, J. Phys. A \textbf{22}, 865 (1989).

\bibitem{LH} G. Lenz and F. Haake, Phys. Rev. Lett. \textbf{67}, 1 (1991). 
  
\bibitem{PR} T. Prosen and M. Robnik, J. Phys. A \textbf{27}, L459 (1994).

\bibitem{Bogo} E.B. Bogomolny, U. Gerland, and C. Schmit, Phys. Rev. E \textbf{59}, R1315 (1999). 

\bibitem{Brody2} T.A. Brody \textit{et al.}, Rev. Mod. Phys. \textbf{53}, 385 (1981). 

\bibitem{nukandchaos} J.M.G. G\'{o}mez \textit{et al.}, Phys. Rep. \textbf{499}, 103 (2011).

\bibitem{Stockmann} H.-J. St\"{o}ckmann, \textit{Quantum Chaos: An Introduction} (Cambridge University Press, Cambridge, 1999).

\bibitem{RobnikReview} M. Robnik, Nonlin. Phenom. Complex Syst. (Minsk) \textbf{23}, 172 (2020). 

\bibitem{Reichl} L.E. Reichl, \textit{The Transition to Chaos: Conservative Classical and Quantum Systems}, 3rd Ed. (Springer, Switzerland, 2021). 

\bibitem{NETs1} J.A. M\'{e}ndez-Berm\'{u}dez \textit{et al.}, Phys. Rev. E \textbf{91}, 032122 (2015). 

\bibitem{NETs2} C.P. Dettmann, O. Georgiou, and G. Knight, Europhys. Lett. \textbf{118}, 18003 (2017).

\bibitem{NETs3} C. Sarkar and S. Jalan, Chaos \textbf{28}, 102101 (2018). 

\bibitem{NETs4} T. Raghav and S. Jalan, Physica A \textbf{586}, 126457 (2022).

\bibitem{fancy1} R. Potestio, F. Caccioli, and P. Vivo, Phys. Rev. Lett. \textbf{103}, 268101 (2009). 

\bibitem{fancy2} E.F.N. Santos, A.L.R. Barbosa, and P.J. Duarte-Neto, Phys. Lett. A \textbf{384}, 126689 (2020).  

\bibitem{recapps1} Z. Saleki, A.J. Majarshin, Y.-A. Luo, and D.-L. Zhang, Phys. Rev. E \textbf{104}, 014116 (2021).

\bibitem{recapps2} M. Abdel-Mageed, A. Salim, W. Osamy, and A.M. Khedr, Adv. Math. Phys. \textbf{2021}, 9956518 (2021). 

\bibitem{recapps3} J.K. Yao, C.A. Johnson, N.P. Mehta, and K.R.A. Hazzard, Phys. Rev. A \textbf{104}, 053311 (2021). 

\bibitem{recapps4} T. Ara\'{u}jo Lima, R.B. do Carmo, K. Terto, and F.M. de Aguiar, Phys. Rev. E \textbf{104}, 064211 (2021).

\bibitem{recapps5} S. Behnia, F. Nemati, and M. Yagoubi-Notash, Eur. Phys. J. Plus \textbf{137}, 347 (2022). 

\bibitem{Guhr} T. Guhr, A. M\"{u}ller-Groeling, and H.A. Weidenm\"{u}ller, Phys. Rep. \textbf{299}, 189 (1998).

\bibitem{JohnandLutz} J.M. Nieminen and L. Muche, Acta Phys. Pol. B \textbf{48}, 765 (2017). 

\bibitem{Anderson2008} G.W. Anderson and O. Zeitouni, Commun. Pure Appl. Math \textbf{61}, 1118 (2008).

\bibitem{Chak2015} A. Chakrabarty, R.S. Hazra, and D. Sarkar, Acta Phys. Pol. B \textbf{46}, 1637 (2015).

\bibitem{Ben2015} G. Ben Arous and A. Guionnet, in \textit{The Oxford Handbook of Random Matrix Theory}, edited by G. Akemann, J. Baik, and P. Di Francesco (Oxford University Press, Oxford, 2015). 

\bibitem{AEK2017} O.H. Ajanki, L. Erd\H{o}s, and T. Kr\"{u}ger, Probab. Theory Relat. Fields \textbf{169}, 667 (2017). 

\bibitem{Chatter2006} S. Chatterjee, Ann. Probab. \textbf{34}, 2061 (2006).

\bibitem{Credner2008} K. Hofmann-Credner and M. Stolz, Electron. Commun. Probab. \textbf{13}, 401 (2008).

\bibitem{EYY2012} L. Erd\H{o}s, H.-T. Yau, and J. Yin, Probab. Theory Relat. Fields \textbf{154}, 341 (2012).

\bibitem{Gotze2015} F. G\"{o}tze, A.A. Naumov, and A.N. Tikhomirov, Theory Probab. Appl. \textbf{59}, 23 (2015). 

\bibitem{Naumov2015} A.A. Naumov, J. Math. Sci. \textbf{204}, 140 (2015). 

\bibitem{Adamsky2016} R. Adamczak, D. Chafa\"{i}, and P. Wolff, Rand. Struct. Alg. \textbf{48}, 454 (2016).

\bibitem{Che2017} Z. Che, Electron. J. Probab. \textbf{22}, 30 (2017). 

\bibitem{AEK2019} O.H. Ajanki, L. Erd\H{o}s, and T. Kr\"{u}ger, Probab. Theory Relat. Fields \textbf{173}, 293 (2019).
  
\bibitem{cW1} L. Peroncelli, G. Grossi, and V. Aquilanti, Molec. Phys. \textbf{102}, 2345 (2004). 

\bibitem{cW2} H. Yang, F. Zhao, and B. Wang, Physica A \textbf{364}, 544 (2006). 
  
\bibitem{cW3} F.W.K. Firk and S.J. Miller, Symmetry \textbf{1}, 64 (2009).

\bibitem{cWs1} J. Sakhr and J.M. Nieminen, Phys. Rev. E \textbf{73}, 036201 (2006).

\bibitem{cWs2} G. Orjubin, E. Richalot, O. Picon, and O. Legrand, IEEE Trans. Electromagn. Compat. \textbf{49}, 762 (2007). 

\bibitem{cWs3} J.-B. Gros \textit{et al.}, Wave Motion \textbf{51}, 664 (2014).

\bibitem{Commy} H. Matsumura, \textit{Commutative Ring Theory} (Cambridge University Press, Cambridge, 1989).

\bibitem{Indys} A.B. Jaiswal, A. Pandey, and R. Prakash, Europhys. Lett. \textbf{127}, 30004 (2019). 

\bibitem{Germys} G. Akemann, A. Mielke, and P. P\"{a}{\ss}ler, Phys. Rev. E \textbf{106}, 014146 (2022).

\bibitem{beyondsemiPs} H. Hern\'{a}ndez-Salda\~{n}a, J. Flores, and
  T.H. Seligman, Phys. Rev. E \textbf{60}, 449 (1999). 

\bibitem{Bogo2} E.B. Bogomolny, U. Gerland, and C. Schmit,
  Eur. Phys. J. B \textbf{19}, 121 (2001).

\bibitem{JainpapsM2} G. Auberson, S.R. Jain, and A. Khare,
  J. Phys. A \textbf{34}, 695 (2001).  
 
\bibitem{Pbills2} J. Wiersig, Phys. Rev. E \textbf{65}, 046217 (2002). 

\bibitem{GrobeHaake} R. Grobe, F. Haake, and H.-J. Sommers,
  Phys. Rev. Lett. \textbf{61}, 1899 (1988).

\bibitem{MPWQCD} H. Markum, R. Pullirsch, T. Wettig, Phys. Rev. Lett.  
\textbf{83}, 484 (1999).

\bibitem{Ginib} J. Ginibre, J. Math. Phys. \textbf{6}, 440 (1965). 

\bibitem{Norman} N.L. Johnson, S. Kotz, and N. Balakrishnan, \textit{Continuous Univariate Distributions: Volume 1}, 2nd Ed. (Wiley, New York, 1994).

\bibitem{BerryAGN} M.V. Berry and P. Shukla, J. Phys. A \textbf{42}, 485102 (2009). 

\bibitem{Karol} K. \.{Z}yczkowski and M. Ku\'{s}, J. Phys. A \textbf{27}, 4235 (1994).

\bibitem{Magy} A.A. Abul-Magd and A.Y. Abul-Magd, Alex. J. Phys. \textbf{1}, 65 (2011). 

\bibitem{Ahmed2003} Z. Ahmed, Phys. Lett. A \textbf{308}, 140 (2003).

\bibitem{Robnik2007} S. Grossmann and M. Robnik, J. Phys. A \textbf{40}, 409 (2007).

\bibitem{KA2017} S. Kumar and Z. Ahmed, Phys. Rev. E \textbf{96}, 022157 (2017). 

\bibitem{FR2021} J. Feinberg and R. Riser, J. Phys.: Conf. Ser. \textbf{2038}, 012009 (2021). 

\bibitem{CRP1} P. Shukla, Phys. Rev. E \textbf{71}, 026226 (2005). 

\bibitem{CRP2} P. Shukla and S. Sadhukhan, J. Phys. A \textbf{48}, 415002 (2015).

\bibitem{Alt2015} J. Alt, J. Math. Phys. \textbf{56}, 103301 (2015). 

\bibitem{CRP3} A. Kuczala and T.O. Sharpee, Phys. Rev. E \textbf{94}, 050101(R) (2016).

\bibitem{CRP4} J.W. Baron, T.J. Jewell, C. Ryder, and T. Galla, Phys. Rev. Lett. \textbf{128}, 120601 (2022).

\bibitem{nonindR1} B.G. Wilson, F. Rogers, and C. Iglesias, Phys. Rev. A \textbf{37}, 2695 (1988).

\bibitem{nonindR2} J. B\"{u}nemann and F. Gebhard, J. Phys.: Condens. Matter \textbf{29}, 165601 (2017).

\bibitem{WRSG} H. Cao and J. Wiersig, Rev. Mod. Phys. \textbf{87}, 61 (2015).

\bibitem{PTG1} V.V. Konotop, J. Yang, and D.A. Zezyulin, Rev. Mod. Phys. \textbf{88}, 035002 (2016).

\bibitem{PTG2} R. El-Ganainy \textit{et al.}, Nat. Phys. \textbf{14}, 11 (2018).

\bibitem{NHP} Y. Ashida, Z. Gong, and M. Ueda, Adv. Phys. \textbf{69}, 249 (2020). 

\bibitem{Poli} C. Poli, B. Dietz, O. Legrand, F. Mortessagne, and A. Richter, Phys. Rev. E \textbf{80}, 035204(R) (2009). 

\bibitem{koreans} J.W. Ryu and S.W. Kim, Chaos \textbf{29}, 043123 (2019).

\bibitem{Wett12} S. Schierenberg, F. Bruckmann, and T. Wettig, Phys. Rev. E \textbf{85}, 061130 (2012). 

\end{thebibliography}
\end{document}